\documentclass[twocolumn]{aastex631}
\usepackage{amsmath,amsfonts,amssymb,graphicx,chngcntr,multirow, float,booktabs, afterpage}
\usepackage{longtable}

\shorttitle{Triple radio flares from TDEs}

\begin{document}

\title{Triple radio flares from tidal disruption events: jet-wind collisions and the discovery of a third radio flare from AT2020vwl}

\author{A. Mummery} 
\thanks{These authors contributed equally to this work.}
\affiliation{School of Natural Sciences, Institute for Advanced Study, 1 Einstein Drive, Princeton, NJ 08540, USA}
\email{amummery@ias.edu}
\author[0000-0003-3441-8299]{A. J. Goodwin}
\thanks{These authors contributed equally to this work.}
\affiliation{International Centre for Radio Astronomy Research – Curtin University, GPO Box U1987, Perth, WA 6845, Australia}
\email{ajgoodwin.astro@gmail.com}  
\author[0000-0003-0528-202X]{C. T. Christy}
\affiliation{Department of Astronomy/Steward Observatory, 933 North Cherry Avenue, Rm. N204, Tucson, AZ 85721-0065, USA}
\author[0000-0002-8297-2473]{K. D. Alexander}
\affiliation{Department of Astronomy/Steward Observatory, 933 North Cherry Avenue, Rm. N204, Tucson, AZ 85721-0065, USA}
\author[0000-0003-4537-3575]{N. Franz}
\affiliation{Department of Astronomy/Steward Observatory, 933 North Cherry Avenue, Rm. N204, Tucson, AZ 85721-0065, USA}

\begin{abstract}
Radio flares from tidal disruption events (TDEs) are observed on two distinct
timescales: prompt (around the time of optical peak) and delayed (hundreds to thousands of days after optical peak). A recent framework put forward to explain this diversity suggests that super-Eddington winds produce prompt flares while jets launched during a disk state transition at  low accretion rates produce delayed flares. A unique observational prediction of this framework is that individual sources showing both types of flares may be expected to show a third flare. This third flare will occur when the (faster) jet catches up with the (slower) wind. The observational appearance of the third flare depends on the system parameters. As the two shock fronts collide the (brighter) jet encounters a change in density, modifying its synchrotron-self-absorption frequency and observed flux. Here we discuss the observational appearance of such a scenario, and then show, in new long-term monitoring radio observations of the TDE AT2020vwl, the discovery of a third radio flare. This flare is coincident with the time at which forward modeling of the first two flares in a wind-jet framework predicts a collision between the two shock fronts. We  make predictions for the long-term radio evolution of two other TDEs showing double radio flares: ASASSN-15oi (which may show a weak flare or flattening of the lightcurve) and AT2024tvd (which is predicted to imminently undergo a third flare). 
\end{abstract}


\keywords{
Accretion (14);
High energy astrophysics (739); 
Supermassive black holes (1663);\\
X-ray transient sources (1852); 
Time domain astronomy (2109)
}


\section{Introduction}
Tidal disruption events (TDEs), in which a star is destroyed by the tidal
forces of a supermassive black hole \citep[e.g.][]{Rees88}, have become powerful probes of accretion and outflow physics around otherwise quiescent black holes. A growing fraction are detected at radio wavelengths \citep[e.g. see][for sample studies]{Alexander20,Cendes24,Goodwin25b}, where synchrotron emission from an outflow shocking the circumnuclear medium (CNM) offers the cleanest diagnostic of the energetics, velocity, and geometry of the ejecta \citep[e.g.][]{Alexander16,Cendes21,Goodwin22,Goodwin23b,Christy24,Golay25}, and of the gas density on sub-parsec scales \citep[e.g.][]{GoodwinMummery26b}.

\begin{figure*}
    \centering
    \includegraphics[width=0.95\linewidth]{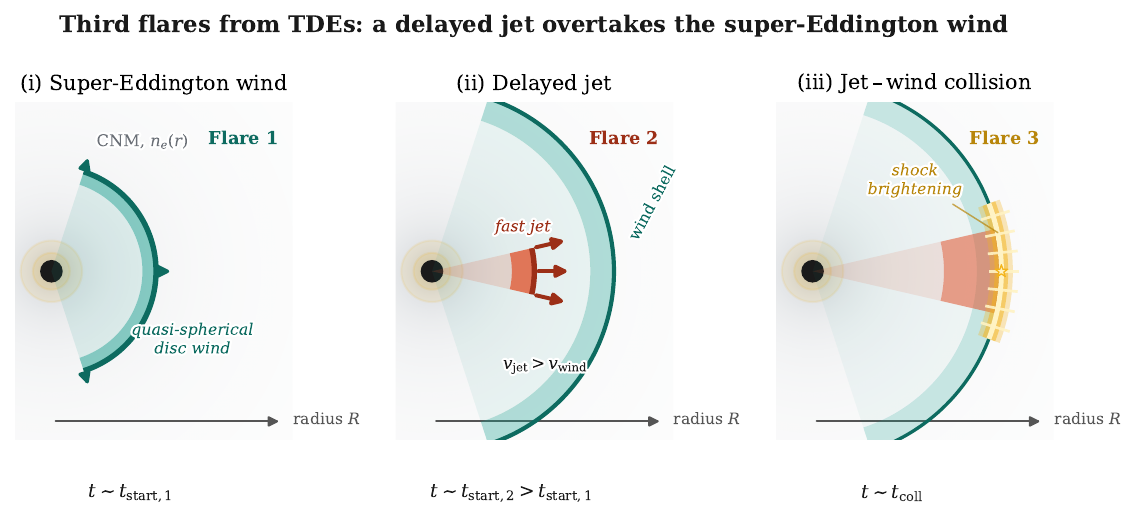}
    \caption{A schematic of the framework put forward in \citealt{GoodwinMummery26} and extended here. Prompt radio flares in tidal disruption events are produced by the interaction of a super-Eddington ($\dot m \gtrsim 1$) wind with the circumnuclear material, which may then be followed at  late times by a state transition jet (when the accretion rate drops through $\dot m < \dot m_{\rm crit} \sim 0.03$). Jets generically move faster than winds (although X-ray binary state transition jets are observed to be $\Gamma < 2$ systems, so not the very relativistic jets of some TDEs), implying that at yet later times the jet will catch up with the wind and encounter a change in density, which is expected to propagate into a change in synchrotron self-absorption frequency and a third radio flare.   }
    \label{fig:schem}
\end{figure*}

Radio bright TDEs display a notable diversity in the timing of the flares observed. Some brighten
promptly, within weeks to a few months of optical peak
\citep[e.g.][]{Alexander16,Cendes21,Goodwin22}, others remain radio-quiet
for months to years before exhibiting a delayed flare
\citep[e.g.][]{Horesh21,Cendes23, Goodwin2025}, while others remain undetected in radio \citep{Alexander20}. Systematic late-time monitoring has shown that delayed radio emission is
common rather than exceptional \citep{Cendes24}, and that some sources
exhibit \emph{multiple} distinct radio flares
\citep[e.g.][]{Horesh21,Cendes23,Goodwin2025,Sfaradi25}. The physical origin of this
behaviour has been debated, with proposed explanations including
freely-expanding sub-relativistic winds \citep[e.g.][]{Alexander16,Goodwin2025}, the delayed launching of a relativistic (bulk $\Gamma \gg 1$) jet \citep[e.g.][]{Teboul23,Lu2024}, the deceleration of off-axis relativistic jets \citep{Matsumoto23}, disk instabilities \citep{Wu26}, disk state transitions \citep{Giannios11, ALexander26, GoodwinMummery26} and non-uniform density structure
in the CNM \citep[e.g.][]{Matsumoto24,Zhuang25}.

A unifying framework was recently proposed in which  radio flares in TDEs correspond to two physically distinct outflows
\citep{GoodwinMummery26}: prompt flares are driven by a super-Eddington
disc wind launched while the (Eddington normalised) accretion rate is high ($\dot m \equiv \dot M_{\rm acc}/\dot M_{\rm Edd} \gtrsim 1$),
whereas delayed flares are produced by a jet launched at a soft-to-hard
state transition, when the accretion rate falls below a critical value
($\dot m \lesssim \dot m_{\rm crit}\sim0.03$). 

We clarify terminology here, as the word ``jet'' means slightly different things in different communities. Black hole X-ray binaries (to which we continually make analogy here) launch two different types of outflows \citep{Fender04}, both called ``jets'', at two different mass accretion rates with different kinematic properties. At high $\dot m$ (as the source moves from the hard to soft state), relativistic (i.e., bulk $\Gamma \sim 5-10$ or more) jets are launched, the analogue of (e.g.,) the jet seen in the relativistic TDE Swift J1644 \citep{Zauderer11,Berger12}. At the low $\dot m$ end, a kinematically distinct type of jet is launched as the hard state is entered \citep[e.g.][]{Fender01}. These ``compact'' jets are slower, typically $\Gamma \sim 1-2$. In the transients community these are sometimes simply called ``outflows'', with ``jet'' reserved for the highly relativistic end of this framework, but as we are seeking to make connections to the state of the accretion disk in this framework, we shall follow the X-ray binary/accretion community terminology. 
Independent of semantics, within this picture the ``wind''
is slow(er) and quasi-spherical and the ``jet'' is fast(er) and likely more collimated, and the two are launched at distinct separated times within the same event. 

The dual outflow framework makes a clean and unavoidable prediction. If a single TDE launches both a wind and a later jet, then because jets generically expand faster than winds, the jet shock front must eventually overtake the
wind-blown shell. When it does, it encounters an abrupt change in the
density -- the dense, swept-up post-shock wind shell -- which raises its
synchrotron self-absorption frequency and self-absorbed flux, producing a
\emph{third} radio flare (or, at minimum, a marked change in radio
properties) with no change in the central engine. A source that has shown
both a prompt and a delayed flare may therefore show a third, at a time
set purely by the dynamics of the first two flares.

In this 
paper we test this third flare prediction. We first overview the wind-jet collision physics and show that the timing of the third flare is an essentially kinematic consequence of the first two flares, while its brightness and spectral appearance depends on more uncertain post-collision shock physics. We then present new radio observations of AT2020vwl obtained as part of a long-term monitoring campaign, revealing a third flare at the time the forward modelling of its first two flares predicts the two shock fronts collide. Finally, we apply the same modelling to the other two double-flare
TDEs with well-constrained lightcurves: ASASSN-15oi, predicted to show at most a weak, late change, and AT2024tvd, predicted to undergo an imminent third flare. 

The layout of this paper is as follows. In section \ref{sec:theory} we discuss a forward model applicable to radio flares produced by both winds and jets, and the likely observational signatures of a jet-wind collision. In section \ref{sec:data_vwl} we model the radio emission from the TDE AT2020vwl. We show that a third radio flare is observed coincident with the time at which forward modeling of the first two flares predicts a collision between the wind and the jet. In section \ref{sec:data_other} we model ASASSN-15oi and AT2024tvd, showing that ASASSN-15oi may show at most weak further radio evolution, while AT2024tvd is expected to imminently show a prominent third flare. We discuss our results in section \ref{sec:discussion} and conclude in \ref{conc}. 

\section{The Physics of Jet-wind collisions}\label{sec:theory}
The argument discussed in the introduction is qualitative but expected to be generic: a fast jet launched after a slower wind must eventually overtake it. To turn this into a quantitative statement we build a forward model of both outflows, fit it to the first two radio flares of TDE sources, and ask when, and with what observable consequence, the two shock fronts meet. We show a sketch of our framework in Figure~\ref{fig:schem}: an early-launched super-Eddington wind sweeps the circumnuclear medium (CNM) into a decelerating shell (flare~1); a faster, delayed jet drives a second shock through the gas left
in the wind's wake (flare~2); and at later times the jet catches up with the wind shell, where the density jump it encounters powers a third flare.

We model each flare as a mass-loaded (possibly) relativistic outflow that shocks its ambient medium and radiates synchrotron emission. The remainder of this section describes the forward model we use for quantitative predictions, summarising in turn the outflow dynamics (\S\ref{sec:dynamics}), the synchrotron light curve (\S\ref{sec:emission}), the collision and its radio
signature (\S\ref{sec:collision}), and the inference procedure
(\S\ref{sec:inference}). Our aim in this work is to convey the physics of the model. A
full derivation of the analytical and numerical methods, including the internal-shock jump conditions, the numerical implementation, etc. is deferred to a companion code paper (Mummery \& Goodwin, in prep.).

\subsection{Relativistic outflow dynamics with continuous energy injection}
\label{sec:dynamics}
Accretion onto a black hole is inefficient in the sense that not all of the material which originates at large radii makes it onto the black hole. Some fraction is inevitably blown out either into a wind, or a jet. We call either outflow mechanism a ``central engine''. 
We assume that the central engine injects mass at a rate $\dot M(t)$ with velocity
$v_{\rm inj}=\beta_{\rm inj}c$ (bulk Lorentz factor $\Gamma_{\rm inj}$) over a
finite window $t\in[t_{\rm start},t_{\rm end}]$. We adopt a simple
power-law injection history,
\begin{equation}
    \dot M(\tau) = \dot M_0\left(1+\tau\right)^{-q},
    \qquad \tau \equiv (t-t_{\rm start})/(1 \, {\rm day}),
    \label{eq:mdot}
\end{equation}
where the index $q$ is free (the rate declines for $q>0$, rises for
$q<0$),  and the normalisation $\dot M_0$ is set by the total
injected mass $M_{\rm inj}=\int_{t_{\rm start}}^{t_{\rm end}}\dot
M\,{\rm d}\tau$. We adopt a time normalisation of one day so that the injection rate is a pure power law for all relevant timescales on which TDEs evolve (which are much longer than one day).  This is the simplest member of a family of injection profiles
 detailed in the code paper, chosen to minimize the dimensionality of the parameter space. The leading forward shock
sweeps up the CNM of density $n_e(r)$ within a solid angle $\Omega$ (i.e., we assume spherical symmetry of the CNM)
\begin{equation}
    M_{\rm sw}(R) = \Omega\, m_p \int_0^R n_e(r)\, r^2\, {\rm d}r ,
    \label{eq:swept}
\end{equation}
and decelerates as it mass-loads. Because the engine continues to
inject mass with velocity $v_{\rm inj}$, trailing ejecta catch the now-slower leading shock at the retarded time
\begin{equation}
    t_{\rm catch}(t) = t - \frac{R(t)}{v_{\rm inj}},
    \label{eq:catch}
\end{equation}
such that an amount of matter 
\begin{equation}
        \Delta M_{\rm caught}(t) = \int_{t_{\rm start}}^{t_{\rm catch}(t)}
    \dot M(t')\,{\rm d}t' ,
\end{equation}
has caught up to the leading edge of the shock at time $t$, which
re-energises the shock front and drives an internal shock which contributes to the leading-shock emission. The shock
Lorentz factor then follows from energy conservation between the
caught-up ejecta and the swept-up shell. The solution of this balance is \citep{Peer2012}
\begin{equation}
    \Gamma = \frac{2(\Gamma_{\rm inj}+\epsilon)}
                  {1+\sqrt{1+4\Gamma_{\rm inj}\epsilon+4\epsilon^2}},
    \qquad
    \epsilon \equiv \frac{M_{\rm sw}}{\Delta M_{\rm caught}},
    \label{eq:gamma}
\end{equation}
where $\Gamma$ is the bulk Lorentz factor of the shocked shell (not to
be confused with the electron Lorentz factors $\gamma$ below),
and 
\begin{equation}
    R(t)=\int_{t_{\rm start}}^t \, \beta c\,{\rm d}t' ,
\end{equation}
meaning that the dynamics of the shock front satisfies a coupled integro-differential equation system which we solve numerically for a given $\dot M(t)$ and $n_e(r)$ profile. 

Equation~(\ref{eq:gamma}) is the  solution of the forward$+$internal
shock energy balance \citep[the adiabatic bridging relation of][]{Peer2012},
and recovers the standard blast-wave solutions in each corner of the
velocity--mass-loading parameter space. For $\epsilon\to0$ (swept-up mass negligible
relative to the injected mass) it gives $\Gamma\to\Gamma_{\rm inj}$ (i.e., free coasting at the injection speed, at any velocity). In the mass-loaded
regime $\epsilon\gg1$ the limit depends on the velocity: for a Newtonian
injection ($\Gamma_{\rm inj}\to1$) it reduces to
$\beta^2\to\beta_{\rm inj}^2\,\Delta M_{\rm caught}/(2M_{\rm sw})\propto
E/M_{\rm sw}$, the energy-conserving Sedov-Taylor solution
(i.e., the solution which gives $R\propto t^{2/(5-k)}$). For a relativistic injection the ordering of $\epsilon$ and $\Gamma_{\rm inj}$ matters. In the intermediate
range $1\ll\epsilon\ll\Gamma_{\rm inj}$ the solution reduces to
$\Gamma^2\to\Gamma_{\rm inj}\,\Delta M_{\rm caught}/M_{\rm sw}\propto
E/M_{\rm sw}$, the relativistic Blandford--McKee solution
\citep{BlandfordMcKee1976}. Once $\epsilon\gtrsim\Gamma_{\rm inj}$ the
flow becomes trans-relativistic and rejoins the Newtonian Sedov-Taylor branch,
eventually stalling ($\beta\to0$) as $\epsilon\to\infty$. Finally, because the engine
injects continuously ($\Delta M_{\rm caught}$ grows with time), the
Newtonian decelerating regime under ongoing injection reproduces
self-similar, Chevalier-type ejecta--CNM interaction dynamics
\citep{Chevalier1982,Chevalier1998}, with the continuous mass injection
playing the role of Chevalier's steep ejecta-density profile (the
energy-injection/refreshed-shock equivalence of \citealt{SariMeszaros2000}). We
retain the relativistic form throughout (as it trivially reproduces the Newtonian limit, but allows for more freedom if the data wants it), although the velocities
inferred for many of our TDEs are only mildly relativistic ($\beta \sim 0.2-0.8$, consistent with the $\Gamma<2$ expected in the hard state jet regime). 

The two outflows propagate into different environments. The wind, expanding first, plows into the pristine circumnuclear gas, which we take to be a Bondi profile fixed by the black-hole mass $M_\bullet$, the Eddington ratio $f_{\rm Edd}$, and the ambient large-radius gas temperature $T_\infty$ \citep[see][for a full derivation of the Bondi profile we use]{GoodwinMummery26b}. The jet,
launched later, does not see this pristine medium. We assume that it propagates into the gas left in the wake of the wind which has already swept up and re-arranged by the earlier outflow. Therefore, rather than impose the Bondi density distribution, we model its ambient density with a free power law $n_e\propto r^{-k}$ with normalisation $n_0$ (which we define as the density at radius $R_0=10^{16}$\,cm). The differing deceleration rates of the two shocks as well as their initial launch speeds thus determine whether and when they collide.

\subsection{Synchrotron emission and the radio light curve}
\label{sec:emission}
At its outward propagating front, each outflow shocks the ambient medium owing to its super-sonic propagation. This shock front both amplifies magnetic fields in the surrounding plasma and also accelerates ambient electrons. Together this produces synchrotron radiation. 

We follow standard synchrotron modeling assumptions throughout. At the shock a fraction, $\varepsilon_e$, of the post-shock energy density
$e'$ is placed in a power-law electron distribution
$N(\gamma)\propto\gamma^{-p}$ above a minimum Lorentz factor $\gamma_m$. A (generally different) fraction, $\varepsilon_B$, is placed into the magnetic field
$B'=\sqrt{8\pi\varepsilon_B e'}$. It is not always possible to accelerate all of the ambient electrons into a non-thermal power law distribution, especially at late times when the shock front may have decelerated.  If the shock front decelerates  such that it has insufficient energy to accelerate all swept up electrons into a non-thermal distribution, then we place only a fraction $f_{\rm DN}$ (chosen to conserve energy) into this non-thermal distribution, with the rest of the electrons placed into a (non-radiating) thermal population. In the literature this is known as the ``deep-Newtonian'' regime, and  we suppress the emitting electron population by the standard ``deep-Newtonian factor'' \citep{Sironi2013}.

The synchrotron emission is self-absorbed below the frequency $\nu_a$
(the frequency at which  the optical depth to self absorption $\tau_\nu=1$) and with characteristic frequency
$\nu_m$ set by $\gamma_m$ and $B'$. We assume that the shock accelerates electrons in a shell of thickness $\Delta R$, where we compute $\Delta R$ by mass conservation (the number of electrons in the post shock density $n_e'$ within $\Delta R$ of the shock front must equal the number of electrons swept up by the shock). 

For the self-absorbed regime
relevant here ($\nu_a\gg\nu_m$) the spectrum is a (smoothly) broken
power law, optically thick ($F_\nu\propto\nu^{5/2}$) below $\nu_a$ and
optically thin ($F_\nu\propto\nu^{-(p-1)/2}$) above it, with peak flux
\begin{equation}F_{\nu_p}=F_{\nu_m}\left(\frac{\nu_a}{\nu_m}\right)^{\frac{1-p}{2}}
    \label{eq:spectrum}
\end{equation}
\citep{Granot2002,BarniolDuran2013}. The observed frequency and flux are
boosted by the standard relativistic Doppler factor $\delta(\theta_{\rm obs})$ for viewing
 angle $\theta_{\rm obs}$. 

The density dependence of $\nu_a$ is central to what follows. Since
$\nu_a$ marks where the self-absorption optical depth reaches unity, and
$\tau_\nu\propto n_e'\,B'^{(p+2)/2}\,\Delta R\,\nu^{-(p+4)/2}$ for a
power-law electron population (where $\Delta R$ is again the thickness of
the shocked emitting shell behind the forward shock) the
self-absorption frequency, which scales as
$\nu_a\propto\left(n_e'\,B'^{(p+2)/2}\,\Delta R\right)^{2/(p+4)}$, grows
with the post-shock density and magnetic field, and hence with the
ambient density into which the shock propagates, since $n_e'\propto n_e$
and $B'\propto n_e^{1/2}\beta$. The self-absorption frequency thus behaves as
a measure of the density into which the shock front enters, a shock that enters denser gas
drives both $\nu_a$ and the self-absorbed peak flux
$F_{\nu_p}\propto\nu_a^{5/2}$ upward, re-brightening the source with no
change in the central engine. This is the mechanism we invoke for the third flare, in which the delayed jet overtakes the dense, swept-up wind
shell.

We include synchrotron cooling, though its effects on two of the three sources we model (ASASSN-15oi and AT2020vwl) are completely negligible. The standard cooling frequency
$\nu_c\propto B'^{-2}t^{-1}$ lies well above their observing band throughout (median $\nu_c\gtrsim 100\times$ the highest observed
frequency across the posterior for these two sources), so the radio-emitting electrons are comfortably in the
slow-cooling regime. For AT2024tvd, which has observations with ALMA at $\sim100$\,GHz frequencies, there is evidence for a cooling break at late times, though only a small number of data points are impacted and this does not dominate our overall fit statistic. 

\subsection{Discussion of the jet--wind collision}\label{sec:collision}
With a forward model for each flare (the wind and the jet), we
can then compute the two radii of the shock fronts
\begin{align}
    R_{\rm wind}(t) &= \int_{t_{\rm flare\,1\,start}}^{t}\,
        \beta_{\rm wind}\, c\, {\rm d}t' ,\\
    R_{\rm jet}(t) &= \int_{t_{\rm flare\,2\,start}}^{t}\,
        \beta_{\rm jet}\, c\, {\rm d}t' ,
\end{align}
where we solve the full dynamical equations (a function of
$\dot M(t)$, $v_{\rm inj}$, $n_e(r)$, etc.) for $\beta(t)$ in both models.
In this model a collision occurs at a time $t_{\rm coll}$ given by definition by
\begin{equation}
    R_{\rm wind}(t_{\rm coll}) = R_{\rm jet}(t_{\rm coll}).
\end{equation}
The question then becomes what is the observable consequence of such a collision. A similar sort of problem (although in very different regime) have been considered in the GRB literature \citep[e.g.][for models of faster outflows catching slower shells]{Rees98, Panaitescu98, Nakar07, Laskar15}. 

The collision drives a new pair of shocks: one that is transmitted into the
wind shell and one that is reflected back into the jet. It is the radio
signature of this freshly shocked gas that creates the third flare.
Because the shocks in these systems are at most mildly relativistic (this is an empirical result from our fits), the
characteristic synchrotron frequency $\nu_m$ lies far below the observing
band, and the spectral peak is set instead by synchrotron
self-absorption. The observable consequence of the collision is therefore
governed by how the self-absorption frequency $\nu_a$ responds to the
encounter. In the model, $\nu_a$ depends on the post-shock electron
density, magnetic field, and emitting width as
\begin{equation}
    \nu_a \propto \left(n_e'\, B'^{(p+2)/2}\, \Delta R\right)^{\frac{2}{p+4}},
    \qquad n_e' \propto n, \quad
    B' \propto n^{1/2}\beta ,
\end{equation}
so that, across the collision, the shift in $\nu_a$ factorises into three
dimensionless quantities,
\begin{equation}
    \frac{\nu_a^{\rm coll}}{\nu_a^{\rm jet}} \;\simeq\;
    \underbrace{D^{\frac{p+6}{2(p+4)}}}_{\text{density contrast}}\;
    \underbrace{\left(\frac{\beta_{\rm rel}}{\beta_{\rm jet}}\right)^{\frac{p+2}{p+4}}}_{\text{velocity contrast}}\;
    \underbrace{\left(\frac{\Delta R_{\rm coll}}{\Delta R_{\rm jet}}\right)^{\frac{2}{p+4}}}_{\text{shell compression}} ,
    \label{eq:nua_factors}
\end{equation}
where 
\begin{equation}
    D \equiv {n'_{\rm wind}\over n_{\rm jet}}, \qquad n'_{\rm wind} = C n_{\rm wind},
\end{equation} 
where $C\equiv \Gamma[(\gamma_{\rm ad} + 1)/(\gamma_{\rm ad} -1 )]$ is the strong shock compression density compression ratio (the jet impacts the denser post-shock material of the wind first), and $D$ represents  the density contrast experienced by the jet  relative to its own pre-collision ambient density. The velocity factor  $\beta_{\rm rel} \equiv (\beta_{\rm jet}-\beta_{\rm wind})/
(1-\beta_{\rm jet}\beta_{\rm wind})$ is the speed of the jet as seen from the perspective of the wind. 

Of the three factors, only the density contrast acts to raise $\nu_a$.
When the jet shock overruns the denser, already shocked wind shell,
$D>1$. The other two factors suppress the change in $\nu_a$. Because both the wind shell and the jet
propagate outward, the relative speed with which the jet catches up to the wind obeys $\beta_{\rm rel}<\beta_{\rm jet}$
by construction, so the transmitted shock advances into the receding shell
more slowly than the jet shock moved relative to its up stream medium, generally weakening
the shocks (although of course in the relatively natural limit $\beta_{\rm jet} \gg \beta_{\rm wind}$ then $\beta_{\rm rel}\approx\beta_{\rm jet}$ and there is no suppression at all). The width factor likewise suppresses the frequency jump as the colliding objects are not
cold shells but extended, already shocked blast waves, and the resulting 
shock likely occurs in a somewhat thinner, strongly compressed layer at the contact
discontinuity, so $\Delta R_{\rm coll}<\Delta R_{\rm jet}$. The
re-brightening is therefore primarily driven by the density jump alone, partially (or possibly largely depending mainly on the relative velocities of the two fronts)
offset by both the reduced relative velocity and the thinner emitting
layer.

We can estimate the density contrast and frequency jump in the following idealized limit, while keeping in mind that this is likely an over simplification.  Imagine that the wind has been propagating for a reasonably long amount of time, such that the density behind it is not much perturbed from the  circumnuclear medium $n_{\rm CNM}$ (i.e., imagine if the density behind the wind has settled down into approximately its previous state). In this limit the jet and the wind are propagating through roughly the same density (just at two different times), and the density contrast is then given entirely by the standard strong shock conditions, i.e.,  $n'_{\rm shock} = \Gamma[(\gamma_{\rm ad} + 1)/(\gamma_{\rm ad} -1 )] n_{\rm CNM} = 4\Gamma n_{\rm CNM}$ for a standard $\gamma_{\rm ad} = 5/3$ gas. If the jet rapidly catches up with the wind $\beta_{\rm  rel}\approx \beta_{\rm jet}\gg\beta_{\rm wind}$, then as the two shock fronts are at the same radius (by definition) and as they have swept up the same density profiles have the same shell widths (as their mass conservation constraints are the same), then we can roughly estimate $\Delta R_{\rm coll} \sim \Delta R_{\rm wind} \sim \Delta R_{\rm jet}$, and so  
\begin{equation}
        \frac{\nu_a^{\rm coll}}{\nu_a^{\rm jet}} \sim \left({\gamma_{\rm ad}+1\over \gamma_{\rm ad}-1}\right)^{\frac{p+6}{2(p+4)}} \sim 2.5 ,
\end{equation}
where the final numerical value is only weakly dependent on $2<p<4$, and we have assumed $\Gamma \sim 1$ by the time the catch up happens. 

We stress that the density contrast can be significantly larger than this, however, if the wind has effectively evacuated the circumnuclear medium behind it, which is perfectly plausible. In addition, of course, if the jet only just catches up to the wind $(\beta_{\rm rel} \approx \beta_{\rm jet}-\beta_{\rm wind}\ll \beta_{\rm jet})$, the self absorption frequency might barely increase at all, or could in principle even drop. 

Carrying the $\nu_a$ jump through to the flux, the optically-thick peak flux
scales as 
\begin{equation}
    F_{\nu_p}\propto R^2\,\nu_a^{5/2}\,B'^{-1/2},
\end{equation}
so it rises with the density jump through $\nu_a$, which beats the increase in $B'$ for all natural $(2<p<4)$ indices, and the flux
observed in a given band follows the self-absorbed spectrum,
\begin{equation}
    F_\nu = F_{\nu_p}\times
    \begin{cases}
        (\nu/\nu_a)^{5/2}, & \nu < \nu_a,\\[2pt]
        (\nu/\nu_a)^{-(p-1)/2}, & \nu > \nu_a.
    \end{cases}
\end{equation}
The re-brightening is therefore frequency dependent rather than uniform. A
band lying below $\nu_a$ remains optically thick, where
$F_\nu\propto R^2 B'^{-1/2}\nu^{5/2}$ is independent of $\nu_a$ and responds
only to the modest change in $B'$, so it is essentially unchanged or fades
slightly (depending on $\beta_{\rm rel}/\beta_{\rm jet}$ vs the density contrast). A band above $\nu_a$ sits on the optically-thin tail, where
$F_\nu\propto F_{\nu_p}\,\nu_a^{(p-1)/2}\,\nu^{-(p-1)/2}$ rises as both the
peak flux and $\nu_a$ increase. 

The fractional enhancement is largest, and essentially constant, at all
frequencies at or above the post-collision self-absorption frequency
$\nu_a^{\rm coll}$. There the source is optically thin both before and after
the collision with the same slope $-(p-1)/2$, so the flux ratio is
independent of frequency,
\begin{equation}
  \frac{F_\nu^{\rm coll}}{F_\nu^{\rm jet}}
  = \frac{F_{\nu_p}^{\rm coll}}{F_{\nu_p}^{\rm jet}}
    \left(\frac{\nu_a^{\rm coll}}{\nu_a^{\rm jet}}\right)^{\frac{p-1}{2}},
  \qquad (\nu \gtrsim \nu_a^{\rm coll}).
\end{equation}
Between
$\nu_a^{\rm jet}$ and $\nu_a^{\rm coll}$ the band lies on the optically-thick
branch after the collision but on the faint thin tail before, so the
enhancement is reduced below the plateau by a factor
$(\nu/\nu_a^{\rm coll})^{(p+4)/2}$, while below $\nu_a^{\rm jet}$ the band is
optically thick throughout and barely responds. The third flare therefore
brightens by the same factor across the entire optically-thin band above
$\nu_a^{\rm coll}$, with that onset frequency moving upward as the collision
becomes more violent.



We emphasise that, although $\nu_m$ and $F_{\nu_m}$ both change across the
collision, in the regime $\nu_a\gg\nu_m$ the observed peak and the flux at
all bands at or above $\nu_a$ are set by the optically-thick source
function, $F_{\nu_p}\propto R^2\nu_a^{5/2}B'^{-1/2}$, and are independent of
$\nu_m$; the characteristic frequency governs only $\nu<\nu_m$, far below
the observing band.

\subsection{Parameter inference}
\label{sec:inference}

In this work, we fit the model described above to the multiband radio data of three tidal disruption events: AT2020vwl \citep{Goodwin2023,Goodwin2025}, ASASSN-15oi \citep{Horesh21,Hajela25} and AT2024tvd \citep{Sfaradi25}, all of which have detailed publicly available lightcurves of two radio flares, available from the cited references or from TDE databases \citep{Franz26}.   Although the jet expands within the wind shell, so that its photons
traverse the wind, the two synchrotron components may be summed provided
$\nu_a^{\rm jet}>\nu_a^{\rm wind}$. The jet's optically-thin emission
($\nu>\nu_a^{\rm jet}$) then escapes through a wind that is itself
optically thin, while the wind is optically thick only at
$\nu<\nu_a^{\rm wind}$, below the jet's self-absorption turnover, where the
jet contributes only its faint self-absorbed tail and the wind dominates.
We verify (post fit) that this ordering holds across the relevant epochs of our fits.

The parameter space available to the forward model is large, but fortunately the data are constraining. Formally, for each flare, one has the following free parameters. The density profile must be specified, leaving as many free parameters as one wishes to use to specify $n_e(r)$ (we discuss our choice later). The mass injection profile is specified by a total injected mass budget $M_{\rm inj}$, a slope $q$ and the start and end times $t_{\rm start}, t_{\rm end}$. The shock microphysics parameters $p, \varepsilon_e, \varepsilon_B$ are assumed constant throughout the evolution, and are formally free. The outflow is specified by two further parameters, a solid angle $\Omega$ and injection velocity $v_{\rm inj}$. The observer is specified by an angle $\theta_{\rm obs}$. 

We sample the posterior with the affine-invariant ensemble sampler
\textsc{emcee} \citep{EMCEE}, adopting a Gaussian likelihood with a fractional systematic floor, $\sigma^2=\sigma_{\rm obs}^2+(f_{\rm sys}F_{\rm model})^2$, with $F_{\rm model}$ the flux of the model at the evaluated time. We allow $f_{\rm sys}$ to vary as part of the fit. The parameter space spanned by the model is large, and to reduce the fitting size we run an initial chain (100 walkers and 5000 steps) with all parameters free. We then use this chain to reduce the parameter space for the final fits, where we fix the
microphysical and geometrical parameters that the data constrain weakly or
degenerately, namely $\varepsilon_e, \varepsilon_B$ (of both flares), the prompt outflow solid angle  $\Omega_1=4\pi$ (appropriate for a wind), the observer inclination
$\cos\theta_{\rm obs}$ (which is generally not important for mildly relativistic outflows), to values consistent with the peak of the (generally extremely degenerate parameter space). We discuss the implications of this choice in later sections.  The properties of the quiescent Bondi profile we also fix. We take an asymptotic gas temperature $T_\infty$ fixed so that the asymptotic gas speed of sound is comparable to the velocity dispersion of the galaxy $a_\infty^2 = kT_\infty/\mu m_p \approx \sigma_\star^2$ (the large scale constraint of the galaxy being in hydrostatic equilibrium). We take $M_\bullet$ at
the median of the independent disk-emission fit
(which varies per object, \citealt{GoodwinMummery26, Guolo26}). We fit, per flare, the injection
velocity, total injected mass, $\dot M$ slope, and on/off times, together with
the density parameters ($f_{\rm Edd}$ for the wind; $n_0,k$ for the jet),
the electron index $p$, and the jet solid angle $\Omega_2$.

\section{AT2020{\lowercase{vwl}}}\label{sec:data_vwl}

\begin{figure*}
    \centering
    \includegraphics[width=0.9\linewidth]{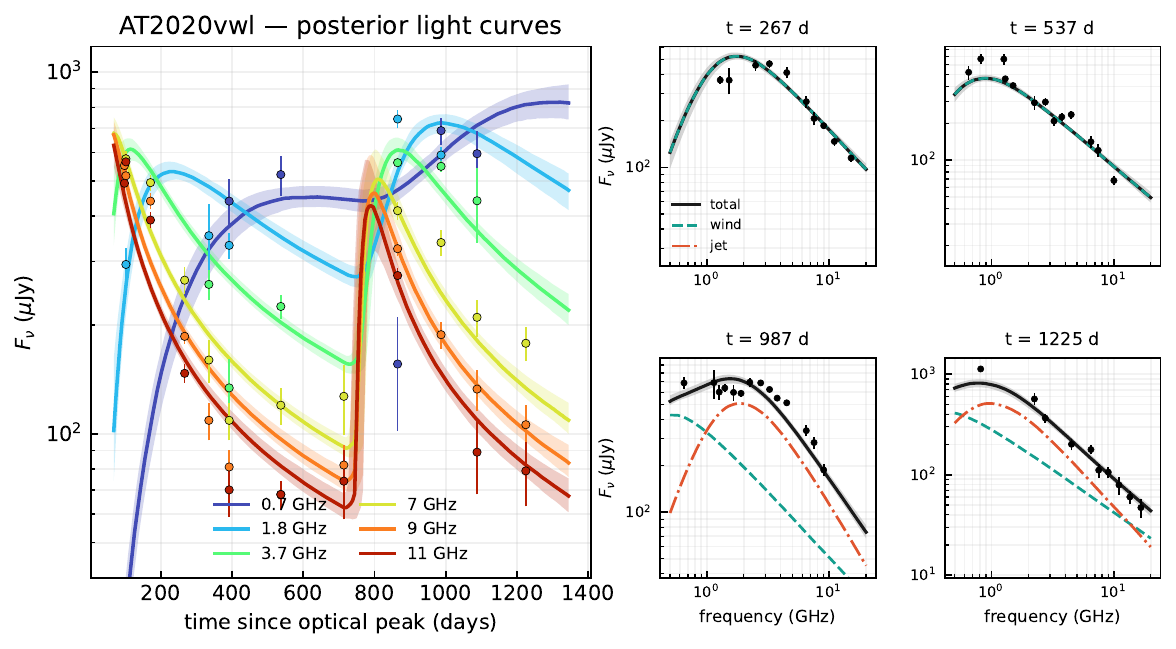}
    \caption{Forward models of the radio light curves of AT2020vwl (first two flares) in a disk wind plus delayed jet framework. On the left we show representative light curves from a range of frequencies (we show only a subset of the 183 data points across 59 frequencies for visible clarity), while on the right we show radio spectra at four distinct epoch, highlighting the transition in spectral dominance from the wind (blue dashed) to jet dominated (red dot dashed) spectral components. The fit is by no means perfect, but clearly captures the bulk evolution of the AT2020vwl system across both frequency and time.  }
    \label{fig:fit_vwl}
\end{figure*}

We begin with modelling AT2020vwl. 
In Figure \ref{fig:fit_vwl} we show the radio afterglow model fit to the first two flares of AT2020vwl (data from \citealt{Goodwin2025}). In the left hand panel we show a set of light curves (we show only a subset of the 183 data points across 59 frequencies for visible clarity), while in the right hand panel we show radio SEDs at four representative times, highlighting the transition in spectral dominance from the wind (blue dashed) to jet dominated (red dot dashed) spectral components.  The fit is by no means perfect (the forward model is of course simplified compared to reality), but clearly captures the bulk evolution of the AT2020vwl system across both frequency and time. It is interesting to note that during the epochs around $\sim 1000$ days (shown by the lower SED panels in Fig. \ref{fig:fit_vwl}) both the wind and the jet are contributing to the emission, explaining the flatter than expected spectral shape below $\sim 2$ GHz, but leading to some difficulty in constraining the spectral index $p$ for this second flare.   

With this forward model, we are able to extract the evolving shock fronts of both the wind and jet (right hand panels of Figure \ref{fig:cross_vwl}). Interestingly, there is a high probability that the two fronts will collide just after the final observational epoch ($P=97\%$ of MCMC samples crossed before day 2000 after optical peak). The posterior distribution of crossing times is relatively strongly peaked around a value $t_{\rm coll} = 1138^{+275}_{-126}$ days from optical peak. The final data point fitted (i.e., the last data point presented in \citealt{Goodwin2025}) was taken at $t = 1225$ days from optical peak. This clearly motivates follow up observations.

\subsection{New radio observations}
We continued monitoring AT2020vwl via our long-term radio monitoring programs with the Karl G. Jansky Very Large Array (VLA) and upgraded Giant Metrewave Radio Telescope (GMRT). In order to test our theoretical prediction for the timing and characteristics of a third flare, in this work we present 5 new epochs of observations of AT2020vwl taken between 2024-07-13 and 2025-12-02 (1334-1718\,d post-optical peak), which can be modelled in addition to the 11 epochs presented in \citet{Goodwin2023} and \citet{Goodwin2025}. Details of the VLA and GMRT radio data reduction are given in Appendix \ref{sec:radio_dat_reduction} and the new radio observations reduced in this work are shown in Table \ref{tab:radio_obs}. 

\subsection{Radio lightcurve}

\begin{figure*}
    \centering
    \includegraphics[width=.49\linewidth]{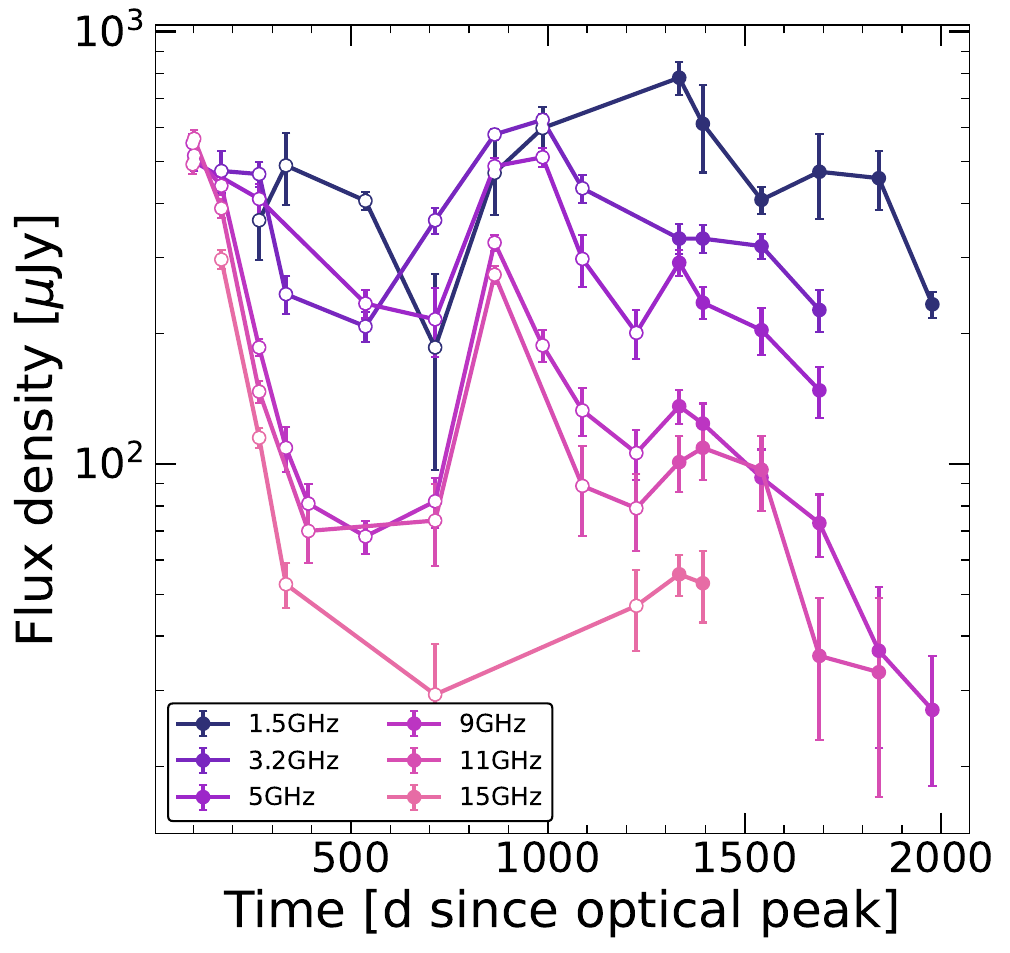}
    \includegraphics[width=.49\linewidth]{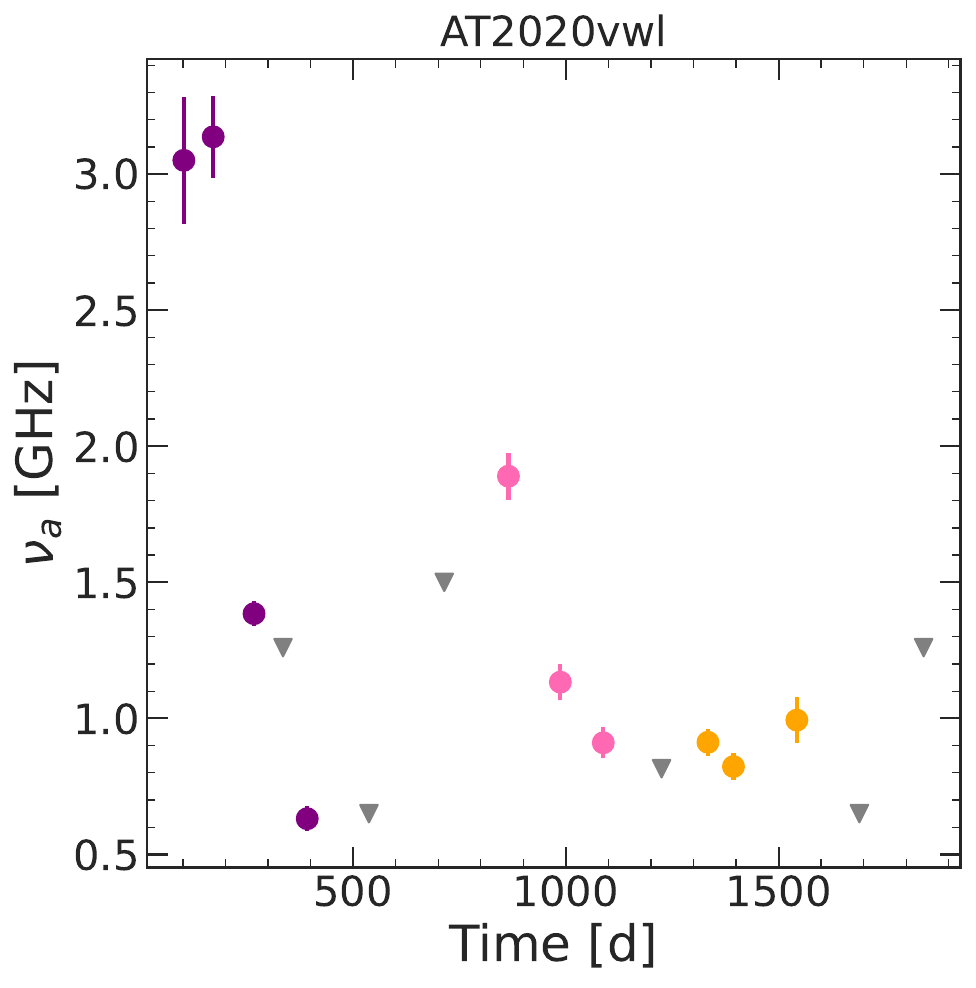}
    \caption{{\bf Left:} The observed radio lightcurves of AT2020vwl from 1.5--15\,GHz. Data from 0--1225\,d are from \citet{Goodwin2023} and \citet{Goodwin2025} (open circles). The new observations presented in this work (closed circles), from 1334--1718\,d post-optical peak, show a distinct re-brightening event observed at all frequencies, which we interpret as a collision between the slower wind and faster delayed jet launched by the TDE. {\bf Right:} The evolution of the synchrotron self-absorption break $\nu_a$, based on epoch-by-epoch spectral fits. The first flare is plotted in purple, the second flare in pink, and the third flare in orange. A subtle increase in $\nu_a$ (observed in the third flare) is exactly what is predicted under the wind-jet collision framework put forward here.  The grey inverted triangles indicate upper-limits on $\nu_a$ where the peak was not well constrained by the spectral fitting procedure.}
    \label{fig:radio_lc}
\end{figure*}

The 1.5--15\,GHz radio lightcurves of AT2020vwl are plotted in Figure \ref{fig:radio_lc}, with the new observations in this work plotted as closed circles. Intriguingly, there is a third radio flare detected commencing in the 1334\,d epoch, exactly within the range of times for the jet-wind collision predicted by the forward model (1138$^{+275}_{-126}$\,d). The magnitude of the third flare is greatest at observing frequencies $>5$\,GHz, and the flare appears to peak at similar times at each frequency band. 

\subsection{Observed evolution of $\nu_a$ during the 3rd flare}\label{sec:spectra}
The jet-wind collision model makes strong predictions about how the radio spectrum (i.e. the location of $\nu_a$) should change at the time of the collision. Here we model the observed radio spectra of AT2020vwl with a simple broken power-law model appropriate for a synchrotron source in which the peak of the spectrum is associated with the self-absorption break \citep{Granot2002},
\begin{equation}
    F_{\nu} = F_{\nu_b} \left[\left(\frac{\nu}{\nu_a}\right)^{-s\beta_1} + \left(\frac{\nu}{\nu_a}\right)^{-s\beta_2} \right]^{-1/s}
\end{equation}
where $\beta_1 = 5/2$, $\beta_2 = (1-p)/2$, $s=5/4 - 9p/50$, and $F_{\nu_b}$ is the flux density normalisation. As in \citet{Goodwin2023} and \citet{Goodwin2025}, we fit the observed spectra using 
\texttt{emcee} \citep{EMCEE}. Notably, given that the temporal modelling in this work allows a constraint on $p$, we fix $p$ in our spectral fits, using $p=2.7$ for the first flare ($t<740$\,d) and $p=3.6$ for the second and third flares. This approach results in slightly different values of $\nu_a$ than presented in \citet{Goodwin2025} where $p$ was allowed to vary freely between epochs, but the constraints on $\nu_a$ are broadly consistent within error. We use log-uniform priors, allowing $10^{-6} < F_{\nu_b} < 10^5$ and $10^{-3} < \nu_a < 10^2$ (where $F_{\nu_b}$ is in $\mu$Jy, and $\nu_a$ in GHz). 

The location of the self-absorption break is is not always well-constrained by the data, in all epochs due it's location and different observational spectral coverage. We determine if $\nu_a$ is constrained by requiring the following criteria is satisfied: $\nu_a$ is constrained to be within the range of frequencies the observations cover ($>0.65$\,GHz) and $\nu_a$ is well-constrained in the MCMC posterior distribution at $>10\sigma$ significance. The GMRT 0.65\,GHz measurements provide crucial constraints on the synchrotron peak in many of the epochs, and the curvature in the band at lower frequencies allows some constraining power on the peak given the fixed spectral model. We report any epochs where $\nu_a$ was completely unconstrained in the fits as upper limits, set to the lowest observed frequency for that spectrum. Plots of the spectral fits for each epoch are provided in the Appendix in Figure \ref{fig:spec_fits}. 

The resulting evolution of the synchrotron self-absorption break is plotted in the right hand panel of Figure \ref{fig:radio_lc}. There is a clear increase in the self-absorption break at the time that the third flare was observed. Comparing $\nu_a$ in the epoch just prior to the onset of the third flare (1225\,d) to the maximum observed $\nu_a$ during the third flare (1543\,d), we find $\nu_{a,\,{\rm coll}}/\nu_{a,\,{\rm jet}}=1.7\pm0.2$. This measurement is formally a lower-limit, as $\nu_{a,\, {\rm jet}}$ would have continued to decrease beyond the measurement at 1225\,d, and the 5 observations between 1225--1841\,d may not have captured the true peak of $\nu_a$.  This increase of $\nu_a$ is entirely consistent with the simplest model of mildly relativistic jet-wind crossings. 

\begin{figure*}
    \centering
    \includegraphics[width=0.99\linewidth]{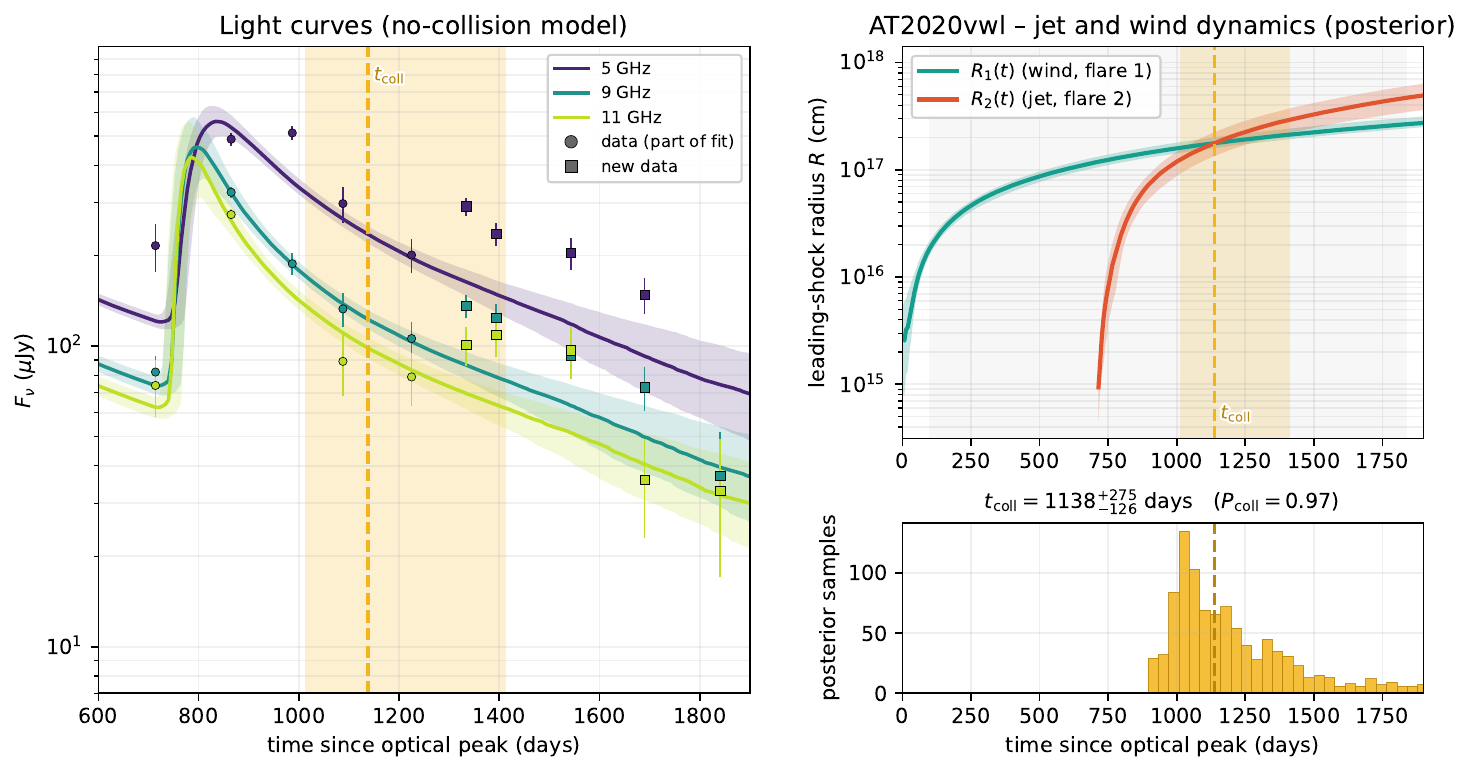}
    \caption{The inferred jet-wind collision time, and the subsequent radio evolution of AT2020vwl. On the left we show the observed and modelled lightcurve of the second radio flare (note the time axis, the first radio flare is not plotted). On the right we show the expansion histories of the two forward shocks in the radio model used to fit the light curves (Fig. \ref{fig:radio_lc}), which (as the jet moves faster than the wind) are inferred to cross at $t_{\rm coll} = 1138^{+275}_{-126}$ days from optical peak. The uncertainty here comes from posterior samples of the MCMC fit to the light curves. At a time consistent with this, AT2020vwl undergoes a third flare observed in multiple bands (left panel shows a subset of the data, all of which is shown in Fig. \ref{fig:radio_lc}).   }
    \label{fig:cross_vwl}
\end{figure*}

To see the increase in observed flux above and beyond the extrapolated forward model fit to the first two flares, see the left hand panel of Figure \ref{fig:cross_vwl}. The deviation from the extrapolated forward model (third flare)  is consistent with occuring at the time at which the dynamical shock front model predicts the jet catches up with the wind.

\section{Other double-flare TDEs: ASASSN-15{\lowercase{oi}} and AT2024{\lowercase{tvd}}}\label{sec:data_other}
\begin{figure*}
    \centering    \includegraphics[width=0.9\linewidth]{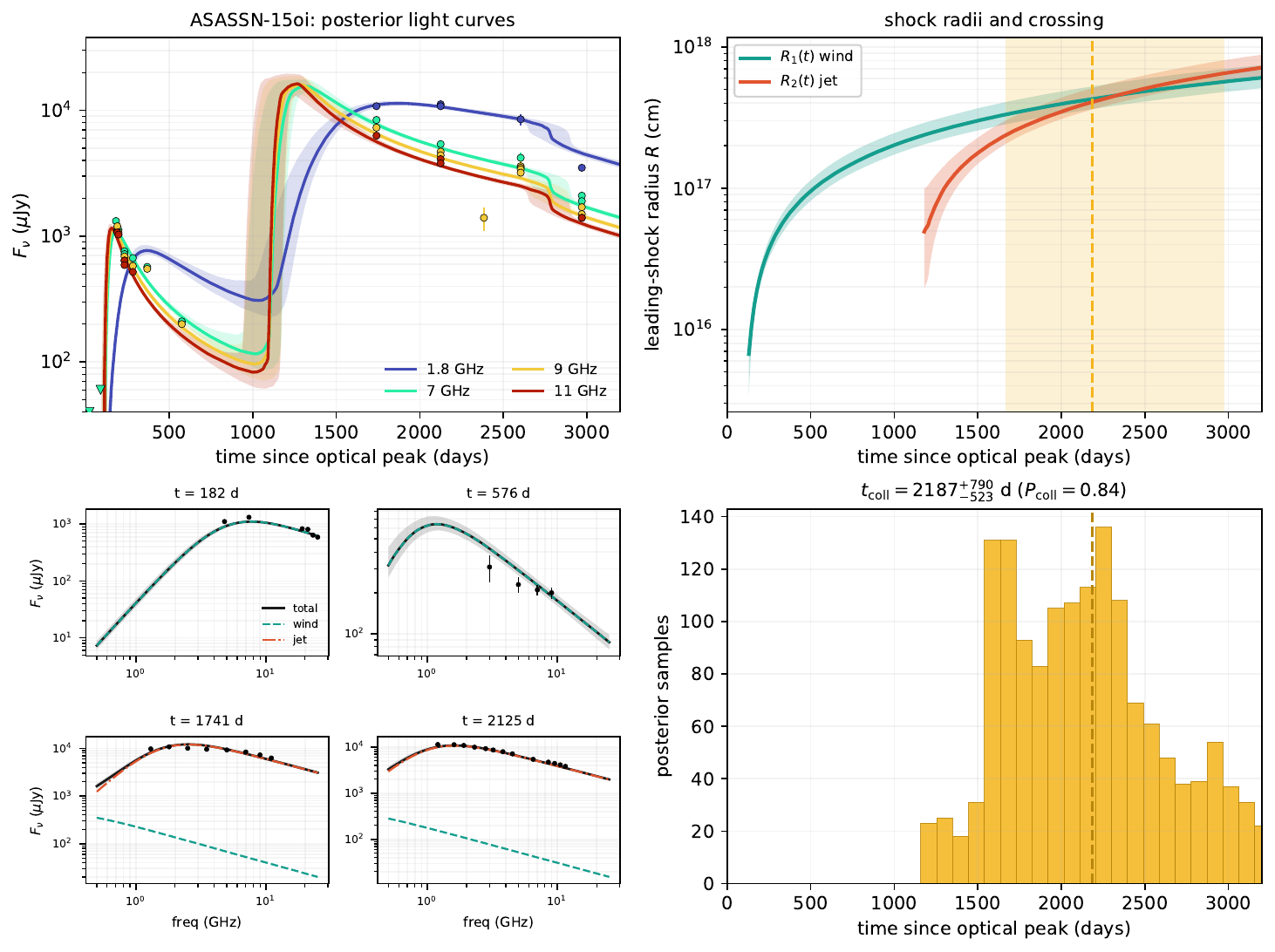}
    \caption{Representative multi wavelength light curves and spectra (left panel; we show only a subset of the 117 data points across 52 frequencies for visible clarity) of ASASSN-15oi, as well as a dynamical model of its two respective shock fronts (right panel). Again, the second outflow is inferred to be more rapidly evolving than the first, leading to an expected collision at a (quite uncertain) late time of $t_{\rm coll} = 2187^{+790}_{-523}$ days after optical peak. While the data coverage is sparse (in both frequency and time) at these late epochs for ASASSN-15oi, and there is no evidence of a clear third flare, the synchrotron spectral peak of ASASSN-15oi is seen to show a break in its spectral evolution at around the time the collision is expected to occur (Fig. \ref{fig:vas_15oi_tvd}).  }
    \label{fig:15oi}
\end{figure*}

\begin{figure*}
    \centering
    \includegraphics[width=0.9\linewidth]{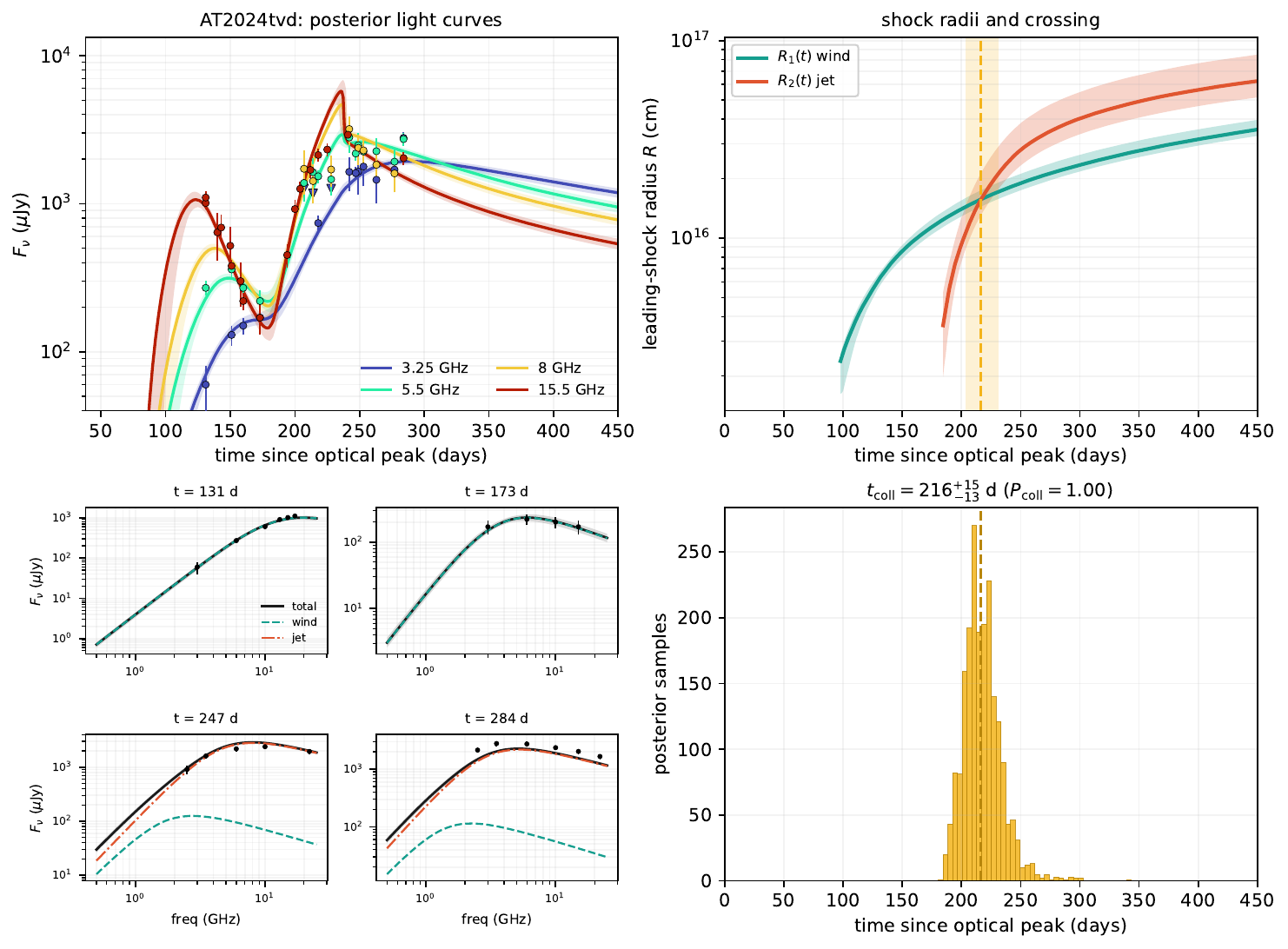}
    \caption{Representative multi wavelength light curves and spectra (left panel; we show only a subset of the 84 data points across 18 frequencies for visible clarity) of AT2024tvd, as well as a dynamical model of its two respective shock fronts (right panel). AT2024tvd showed a very rapid second flare, which is inferred to be moving quickly $v_{\rm inj} > 0.3c$. This leads to an expected collision at a (well constrained in a formal statistical sense)  time of $t_{\rm coll} = 216^{+15}_{-13}$ days after optical peak. We found, however, that the collision time is dominated by systematic uncertainty in the modeling which is enhanced owing to the rapid onset of both flares.  There is tentative evidence for the beginnings of a radio flare at the final epoch of observations ($t = 284$ days post optical peak), and further radio follow up will either confirm or deny the predictions of this framework.  }
    \label{fig:24tvd}
\end{figure*}

\begin{table}
\centering
\caption{Second-outflow (jet) velocity and injected mass, median and
16--84\% intervals.\label{tab:jet}}
\begin{tabular}{lcc}
\hline\hline
Source & $v_{\rm inj,2}\ (c)$ & $M_{\rm inj,2}\ (M_\odot)$ \\
\hline
AT2020vwl   & $0.20^{+0.08}_{-0.06}$ & $(2.7^{+3.7}_{-2.1})\times10^{-2}$ \\
ASASSN-15oi & $0.38^{+0.08}_{-0.08}$ & $(6.2^{+3.6}_{-1.9})\times10^{-3}$ \\
AT2024tvd   & $0.37^{+0.07}_{-0.07}$ & $(1.4^{+1.7}_{-0.8})\times10^{-4}$ \\
\hline
\label{tab:1}
\end{tabular}
\end{table}

The confirmation of a third flare from AT2020vwl means two other double flare sources, ASASSN-15oi and AT2024tvd, warrant study.  We focus on these two sources in this work as they have the most prominent second flares of the known population, we discuss other sources later.  To model these two sources, we take the radio data from \citet{Hajela25} and \citet{Sfaradi25}, and use an identical model setup to that used for AT2020vwl above. 

We begin with ASASSN-15oi. We show the results of our forward modeling for the radio light curves in the upper left panel of Figure \ref{fig:15oi}, with radio SEDs shown in the lower left panel. We show only a representative subset of the 117 data points across 52 frequencies for visible clarity. Once again, the fit is by no means perfect (which would be surprising), but we believe it is capturing the broadband spectral and temporal evolution (ASASSN-15oi is best fit with the engine shutting off just before the final observation, as can be seen by the smooth drop in the radio model at very late times). 

Taking the forward shock propagation of both flares, one extrapolates to a (more uncertain) collision time of $t_{\rm coll} = 2169^{+785}_{-495}$ days post optical peak ($P=86\%$ of MCMC samples underwent a collision before day $+3500$). Perhaps more notably, the jet shock front only just catches up to the wind front (as can be seen by the relative gradients of the two $R(t)$ curves near collision), which predicts a much less pronounced change in radio properties at the intersection of these two shells (to be discussed in more detail in a later section). 

While the light curve of ASASSN-15oi is sparsely sampled, there is tentative evidence for a modest jump in synchrotron self absorption frequency at the latest times (Figure \ref{fig:vas_15oi_tvd}), which could be interpreted as a density increase from a jet catching up to a wind shell. The data is, however, of insufficient coverage to robustly test such a claim.  

We also forward model the evolution of the two flares of AT2024tvd. We note at the outset that AT2024tvd shows two radio flares with peaks separated by only $\sim 100$ days. Unlike ASASSN-15oi and AT2020vwl (with well separated peaks) one should immediately be more concerned with model systematics in such a condensed radio light curve. We model AT2024tvd in an identical fashion to both ASASSN-15oi and AT2020vwl, and discuss systematics in the timing in a   subsection to follow.

We show the evolving radio light curves of AT2024tvd and the forward model in Figure \ref{fig:24tvd}. Again, the forward model is not perfect, but reproduces the bulk of the temporal and spectral evolution of the system. We note that the data favors a delayed start time to flare 1 ($t_{\rm start, \, 1} \approx 80$ d), when the disk is consistent with a super-Eddington accretion rate \citep[this is roughly the time of X-ray peak][]{Guolo26}, but that a quite different solution (invoking free-free absorption in the ambient medium) was found  by \cite{Sfaradi25}, which instead favored an outflow launched at $t\approx 0$ (highlighting immediately a possible systematic uncertainty). 

The forward model predicts that the second outflow moves significantly quicker than the first, and almost immediately catches up with the wind which only had a $\sim 100$ day head start. The collision time is formally well constrained, and every MCMC sample ($P=100\%$) undergoes a collision within  $t=350$ days from optical peak. The uncertainty in the true collision time is dominated by systematic modeling uncertainties, as we discuss later, but the inevitability of a collision in AT2024tvd is reproduced with a wide range of forward model choices. There is a slight jump in the flux (across multiple bands) in the final observation of AT2024tvd ($t=+284$ d), which may be suggestive of a third flare rise. Future radio observations of AT2024tvd will be very interesting. 

More generally, the reader may be surprised that the radio timescales of AT2024tvd are so much shorter than both ASASSN-15oi and AT2020vwl. This, we believe, is genuinely surprising. We believe it unlikely to be causally related to the off-nuclear location of the source, which is presumably just a coincidence.   We note, however, that multi-wavelength evidence suggests that AT2024tvd appears to show qualitatively  similar evolution to ASASSN-15oi  in all wavebands, simply on a much accelerated timescale (i.e., the accelerated evolution is not limited to radio wavelengths). 
ASASSN-15oi was spectrally soft \citep{Holoien16}, and likely obscured \citep{Guolo2024} at early times, before transitioning into a much harder X-ray state at late times \citep{Hajela25,Guolo25c}, after the second radio flare is constrained  to have been launched. 
Note that AT2024tvd is consistent with accreting at  super Eddington rates (just, $\dot m = 1.5^{+0.4}_{-0.3}$) at the time of the first flare launch \citep{Guolo26}, and has a prominent X-ray hard corona in deep XMM spectra during the second radio flare \citep{Yao25,Sfaradi25,Guolo26}, suggesting a disk state-transition powers the second radio flare.

\section{Discussion}\label{sec:discussion}
In this work, we have presented a theoretical framework that predicts the presence of third radio flares in some TDEs. We confirm the detection of a third radio flare in long-term radio monitoring observations of the TDE AT2020vwl, with characteristics consistent with the theoretical predictions of a wind-jet collision. However, there are other scenarios proposed in the literature (to explain second radio flares) that may also provide an explanation for the third flare observed in the radio lightcurve of AT2020vwl. In the following section we discuss some of these alternative scenarios in the context of the observed properties of the third flare from AT2020vwl. We then discuss systematic uncertainty in our collision time inference (which primarily impacts AT2024tvd), and finally provide an overview of the conditions required to power an observable third radio flare in TDEs. 

\subsection{Other potential explanations for a third radio flare}
\subsubsection{Outflow passing through the Bondi radius}
\citet{Matsumoto24} put forward the following model for delayed radio flares in TDEs: if the initial outflow is moving outward through a Bondi sphere, then as the flow reaches the Bondi radius it will encounter a change in density profile (with the density profile flattening). This then produces an observable signature somewhat like what we describe in this work: the density increase (from what was asymptotically expected) leads to a $\nu_a$ evolutionary change, and a change in observed radio properties (a flare). It is possible that such a process could produce a third flare if there were multiple outflows (to explain the first two flares), and one reaches the Bondi radius at some later time. 

We find three main problems with this scenario when applied to AT2020vwl. The first is that a true Bondi sphere has a rather gradual change in density profile, typically over a decade or more in radius. This is not conducive to producing a radio ``flare'', but more a subtle change in decay. \citet{Matsumoto24} model their density profile with $n_e \propto (r^{-k} + 1)$, where the radius normalised by Bondi is $r\equiv R/R_B$, and the profile had a free $k$ ($k$ is equal to 3/2 for a classical Bondi sphere, \citealt{Bondi52}). This is an abrupt change in density, whereas the Bondi sphere is actually much better approximated by $n_e \propto (1 + 1/r)^{3/2}$, which is significantly smoother. Abrupt changes in density lead to abrupt changes in radio light curves, with the corollary that smoother changes produce smoother light curves.  Indeed, we are unable to produce any repeat flaring for any set of Bondi parameters when the actual Bondi sphere solution is used. This is consistent with the results of \citet{Matsumoto24}, who used $k = 2.5-4.5$ in their density profiles to produce a detectable rise in radio emission. 

Second, the Bondi radius needs to be sufficiently close to the central black hole so as to get a delayed rise that happens as quickly as $t \sim 600$ days. An outflow moving at $\beta = 0.1$ for 600 days reaches $R \approx 10^{17}$ cm. For this distance to be the Bondi radius, this requires a gas temperature $T_\infty = 2\mu GM_\bullet m_p/k_B R_B \approx 10^7$ K $(M_\bullet/10^6 M_\odot)$, which is very hot.  To see that this is problematic, we can compute the speed of sound of such a gas $a_\infty \approx (k_BT_\infty/\mu m_p)^{1/2} \approx 500$ km/s $(M_\bullet/10^6 M_\odot)^{1/2}$. While formally any temperature can satisfy the asymptotic Bondi equation for an isolated black hole, black hole accretion flows in galactic centers should satisfy dynamical stability on large (galactic) scales. The velocity dispersion of the galaxy (typically $\sigma_\star \approx 100$ km/s for a TDE host) sets the velocity scale for which objects stay bound to the galaxy itself. If the Bondi sphere really was at $\sim 10^7$ K on large scales, it would be blown immediately out of the galaxy as a wind. There are models for Sgr A* which have $T\sim 10^7-10^8$ K gas near the center (e.g., the \citealt{Quataert04} model), but these are {\it not} Bondi spheres (despite having $n_e \propto R^{-3/2}$ on short scales) exactly because the inner galaxy is outflowing in such a wind. These models predict a {\it steepening} of the density profile beyond the naive Bondi radius one would get from the temperature, not a flattening (see also \citealt{Generozov15, Generozov17, Yalinewich18}). 

Finally, it is not clear which outflow would be traversing the Bondi radius and how that relates to the third flare. Presumably the second (brighter) central ejection would have to be the one that reached the Bondi radius to power a third flare, but then it is not clear why the first ejection did not also show some sort of change when it reached the same radius. 
We therefore find it unlikely that any outflow observed would reach the Bondi radius on the timescales required to power the third flare observed for AT2020vwl.  

\subsubsection{Collisions with a (series of) molecular cloud(s)}
A related (in spirit) model is that of \citet{Zhuang25}, whereby changes in radio emission properties of TDEs at late times relate to a single shock-front encountering (multiple) dense clouds of material in the CNM. Again, the synchrotron emission change is similar to that in our framework (density change leads to $\nu_a$ increase and frequency-dependent radio brightening). 

Of course, one can explain any arbitrarily complicated radio light curve with an arbitrarily complicated CNM, with clouds of varying size, density and distance from the center being encountered by a single outflow. We cannot rule out such a model from radio data (indeed we do not think it is possible to rule out this model), except on fine tuning grounds. 

For such a clumpy cloud based model to explain all of the observations, one would need to invoke a coincidence of why dense clouds are reached just as (i) the X-ray spectra of TDE disks change from soft/thermal states to having power law/coronal components (both ASASSN-15oi and AT2024tvd show such a transition coincident with the second flare launching \citep{Hajela25,Guolo26}), (ii) when independent disk modeling finds a cluster of disk mass accretion rates around a suspicious value $\dot m \approx 0.03$ \citep{GoodwinMummery26}, and (iii) why another dense cloud was reached just when forward modeling of the first two flares of AT2020vwl predict a collision. 

\subsubsection{A newly launched outflow}
A trivial explanation for a third flare is of course always a third outflow. We disfavor such a model for the following reasons. 

Firstly, it is not obvious what disk physics would launch such a third distinct outflow.  Super-Eddington winds and state-transition compact jets are launched in X-ray binaries, but one does not have a natural third outflow mechanism from the disk once the accretion rate has dropped to the low $\lesssim2\%$ Eddington regime, as is likely the case for AT2020vwl \citep[][relativistic jets are discussed shortly]{GoodwinMummery26}. 

An alternative disk mechanism has been proposed in recent models \citep{Wu26}, which invoke the radiation-pressure $\alpha$-viscosity disk instability \citep{LightmanEardley74} to drive  a series of outflows from TDE systems. In the standard \citet{SS73} picture, the turbulent stress that transports angular momentum is represented by a scalar $\alpha$-prescription (an ad-hoc phenomenological proxy for MHD turbulence) taken to scale with the total (gas plus radiation) pressure. Under this assumption the disk becomes both thermally and viscously unstable once radiation pressure dominates the inner region, which occurs above a luminosity of a few percent of Eddington. The instability should manifest as limit-cycle luminosity oscillations in all accreting black hole and neutron star systems above this critical rate.

Observationally, such behavior is vanishingly rare. The high/soft state of black hole X-ray binaries spans the predicted unstable regime ($0.01\,L_{\rm Edd} \lesssim L \lesssim 0.5\,L_{\rm Edd}$), yet these disks show very little variability. Indeed, the high/soft (radiation-pressure-dominated) state is the least variable of all black hole accretion states \citep[e.g.,][]{Belloni05}. Accreting stellar-mass black hole sources cross the critical luminosity both on their rise to peak and on their decline to quiescence without displaying limit-cycle behavior in either direction. The radiation-pressure instability is sometimes associated with the ``heartbeat'' variability state of GRS~1915$+$105. This association is, however, with a single variability class (the ``$\rho$'' class)  of the $12$ cataloged variability states of GRS~1915$+$105  \citep{Belloni00}, with no  mechanism proposed for the remaining 11 classes. This remains one variability state of one exceptional, near- to super-Eddington source, with the great majority of systems crossing the same regime and show no such behavior at all.

As this is supposed to be a scale invariant instability (only caring about the ratio $P_{\rm rad}/P_{\rm gas}$ and not the black hole or disk mass), and there is little evidence for its existence in X-ray binary disks, it is perhaps unsurprising that TDE disks also show no evidence for these limit-cycle instabilities in the optical/UV \citep{Tsuna26}, or more relevantly in the X-rays \citep{Guolo24, Chakraborty26}. In the X-rays any limit-cycle instability should be very obvious, as the Wien-tail nature of TDE X-ray emission exponentially enhances any temperature variability \citep[][which a heating-cooling limit cycle should produce]{MummeryTurner24, Mummery25b}. 

This long-standing scarcity of evidence for this limit cycle is often taken to argue against the total-pressure stress prescription, and has motivated a range of alternatives \citep[see the classic review by][and, more recently, the review of \citealt{Blaes25}]{Done07}. The more fundamental point is that a scalar $\alpha$-closure of any form (whether the stress scales with total, gas, or the geometric-mean pressure) is a substitute for the underlying MHD turbulence, and the instability is a property of one such closure rather than of the disk itself. Consistent with this, the instability is absent in radiation GRMHD simulations that model MHD turbulence directly while also solving for the full radiative stress-energy tensor \citep{Zhang25}. Near-Eddington disks settle into one of two stable, magnetically-mediated configurations rather than undergoing limit-cycles \citep{Zhang26b}, and super-Eddington flows likewise remain stable as thick, turbulent, radiation-supported flows \citep{Zhang26a}. These simulations do not establish that the instability can never arise, but they show that when radiation and magnetic fields are treated self-consistently, the disk is stabilized across all regimes explored to date.

Given the general observational absence of the predicted limit-cycle signatures across black hole X-ray binaries (and X-ray observations of TDEs), and the absence of the instability in first-principles simulations that include radiation and magnetic fields, we regard it as an unlikely explanation for the outflows considered here. Further, as the radiation pressure instability is fundamentally a high accretion rate (in Eddington units) instability \citep{LightmanEardley74, Piro25} it is unlikely to be active at very late times in AT2020vwl, where the Eddington-normalised accretion rate at $\sim 600$ days was already $\sim {\cal O}(0.01)$ \citep{Goodwin25, GoodwinMummery26} from which it will have continued to drop.

One can invoke a third non-disk method of launching material, as in the recent model for delayed radio flares from disk-EMRI collisions of \citet{Linial2026}. However, there are two problems, related to velocity scales, with the origin of the second/third flare being this disk-EMRI collision scenario. First, if one wants the third flare to be produced by the second shell (launched from the collision) catching up with the wind, then one needs a faster second shell than the wind. This is the opposite velocity ordering of the model of \citet{Linial2026}. Indeed, winds are launched at velocities roughly equal to the escape velocity at the super-Eddington trapping radius $R_{\rm tr}$ \citep{Begelman79}, which nearly always works out to be $v_1\sim {\cal O}(0.1)c$ (a result entirely consistent with our forward modeling). By construction, the disk-EMRI collision outflow would be set by the escape speed at the collision radius, which must be smaller than this (i.e., $R_{\rm orb}>R_{\rm tr}$). To see this, note that one does not want a stellar EMRI which collides with the disk which has already been tidally disrupted, and so one must place the EMRI at large radii ($R_{\rm orb}\gg r_T \sim 50R_g$ for a canonical TDE). As the escape speed drops as $\sim c(R_g/r)^{1/2}$, an EMRI which is safe from disruption must be moving more slowly $v_w\sim 0.01-0.05 c$, and therefore the outflow never catches up to the wind. Even if one relaxes the requirement of explaining the third flare (i.e., one invokes a super-Eddington wind, then a collision, then a state transition jet) the velocity constraint is also the origin of the second reason this model does not work for the second flares in our sample. The outflows needed to reproduce the observed data of the second flares for ASASSN-15oi, AT2020vwl and AT2024tvd all have effectively no posterior density below $v_2<0.1c$ (see Table \ref{tab:1}, and corner plots in Appendix). If these second outflows were launched from collisions then the orbital radius would have to be inside the tidal radius of a solar type star around a $10^6 M_\odot$ black hole. 

One could invoke the ordering: (i) super-Eddington wind, (ii) compact state transition jet, then (iii) EMRI-disk collision induced outflow. This runs into fine tuning arguments of why the EMRI-disk outflow switched on exactly when the first two flares were predicted to meet, and why it resulted only in a subtle increase in the self-absorption frequency. 

Indeed, this subtle increase in the self-absorption frequency has an important physical implication which argues against a third distinct outflow being launched from AT2020vwl. This is because standard synchrotron modeling allows us to constrain the radial scale of the shock front from the observed radio spectrum peak flux density ($F_{\mathrm{\nu,a}}$), and the frequency at which the spectrum peaks ($\nu_{\mathrm{a}}$), following the derivation in \citet{BarniolDuran2013}. Schematically, the radius is given by
\begin{equation}\label{eq:R_eq}
R_{\rm shock}(t) 
\propto F_{\mathrm{\nu,a}}^{\frac{6+p}{13+2p}}\nu_a^{-1} ,
\end{equation}
where the proportionality suppresses various micro-physical parameters ($\varepsilon_e, \varepsilon_B$, etc.), which impact the absolute scale of the  radius, but not its temporal evolution (provided they stay roughly constant with time at the shock front). The full expression and discussion is given in Appendix \ref{app:spec}. 

\begin{figure}
    \centering
    \includegraphics[width=0.95\linewidth]{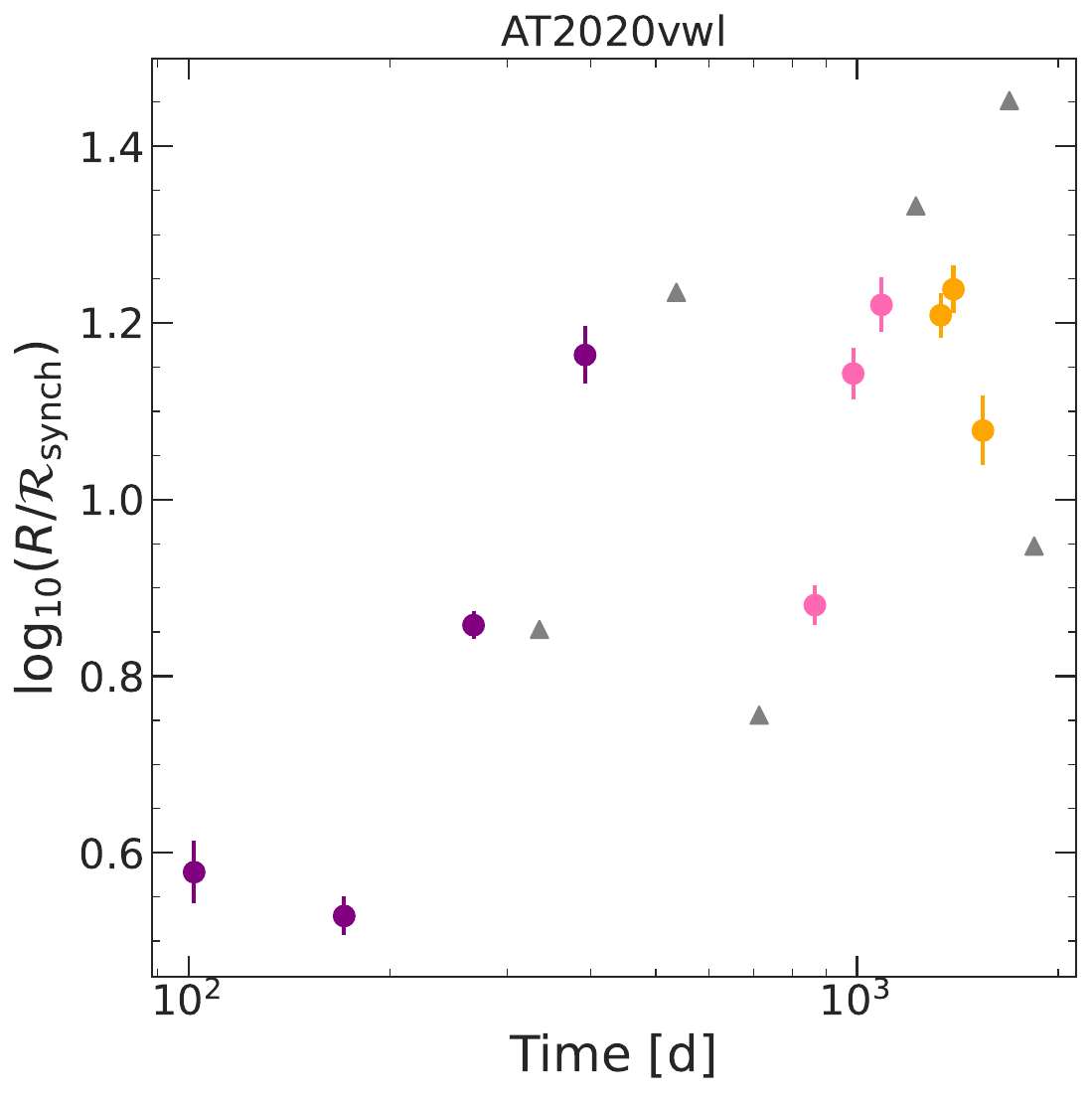}
    \caption{The inferred equipartition radius over time for each of the three flares of AT2020vwl (normalised by the canonical ${\cal R_{\rm{synch}}}$ as the absolute values are not interesting, only the relative values; see Appendix \ref{app:spec} for definition). The transition from flare two (pink) to flare three (orange) is consistent with no observed fall in shock front radius. This is in stark contrast with the radius transition between flares one (purple) and two (pink). This result is entirely consistent with the second shock running over a density contrast, rather than the launching of a third distinct outflow. }
    \label{fig:radius}
\end{figure}

In Figure \ref{fig:radius} we see that, unlike  for the transition between flares one and two, the synchrotron shock radius is broadly continuous across the rise to a third flare. This result is entirely consistent with the second shock running over a density contrast (i.e., catching up with the shocked wind shell), rather than the launching of a third distinct outflow. A third fresh outflow would instead be expected to show a strong discontinuity in the radius evolution, consistent with what is observed between the first two flares.

This lack of radial size evolution also argues against other possible radio-generating outflow orderings. For example, one could argue that the unbound debris stream from the disrupted star might produce the first radio flare, followed by (e.g.,) a super-Eddington wind and then a delayed jet. We disfavor this explanation based on radial sizes, but also the fact that the accretion disk in AT2020vwl is inferred (based on multi-wavelength modeling) to be accreting at only a few percent of Eddington at the $\sim 600$ day window when the second outflow was launched \citep{Goodwin2025, GoodwinMummery26}. 

\subsubsection{A repeating partial TDE}
A possible explanation for multiple $(N>2)$ flaring events in radio could be a repeated partial tidal disruption event, which undergoes one (or both) of a super-Eddington wind/state transition jet multiple times (as the disk is refreshed). This is a known phenomenology in TDEs, with eRASSt J045650.3$-$203750 \citep{Liu23, Liu24} showing a repeated cycling through soft (observed to be radio faint) and hard (observed to be radio bright) X-ray states. This particular repeated partial TDE is of course consistent with our framework for the first two flares of AT2020vwl. We exclude a repeated partial TDE explanation for AT2020vwl based on its optical/UV light curve, which shows only one flare \citep{Goodwin2025}. 

\subsubsection{An off-axis relativistic jet}
A relativistic jet is only observable at angles not limited by beaming $\theta \lesssim 1/\Gamma(t)$, so if such a flow is launched and observed at $\theta > 1/\Gamma_{\rm start}$, the flow must decelerate before it can be observed. In principle (although we find such an idea dubious) one could invoke the following three distinct outflows being launched (i) a relativistic jet observed off axis, (ii) a super-Eddington wind (or other prompt outflow) and then (iii) a state transition (mildly relativistic) compact jet. One could imagine fine tuning various angles and velocities so as to observe three distinct radio rises. We have not attempted to model such a fine tuning, and we deem it very unlikely that such a model could be forced to reproduce only a subtle rise in the synchrotron self absorption frequency at the time of the third flare. This is for exactly the same reason as is discussed above and in Figure \ref{fig:radius}, the subtle change in $\nu_a$ implies an almost continuous radius evolution of the dominant shock front, something which is almost impossible to engineer with two spatially distinct emitting regions coming into view. 

\newpage
\subsection{Systematics in the predicted collision times}
The collision time is defined kinematically, $R_{\rm wind}(t_{\rm coll})=
R_{\rm jet}(t_{\rm coll})$, and therefore depends only on the inferred
velocity histories of the two outflows. The credible intervals reported above
are, however, purely \emph{statistical} --- they propagate the
data-constrained posterior of each forward model --- but a number of
\emph{systematic} effects act in addition to them. We discuss these here but emphasize that numerical experimentation with various of these effects leads to $\Delta t \sim {\cal O}(100)$ day shifts in the collision time. While this is not particularly important for AT2020vwl and ASASSN-15oi, this is an $\sim {\cal O}(1)$ correction for AT2024tvd which seems to evolve on an accelerated timescale compared to other TDEs.

The radius histories follow from the swept-up mass,
$M_{\rm sw}(R)=\Omega\,m_p\int_0^R n_e(r)\,r^2\,{\rm d}r$, so the assumed
circumnuclear density profile propagates directly into $R_{\rm wind}(t)$,
$R_{\rm jet}(t)$, and hence into the crossing. The principal systematic
here  is the density idealization: we represent the wind's ambient medium
by an analytic, spherically-symmetric, steady-state Bondi atmosphere (and
the jet's by a smooth power law to account for this wind perturbing the background), whereas the real circumnuclear gas around
a black hole is unlikely to resemble a perfectly ideal Bondi sphere. It is more plausibly clumpy, multiphase, aspherical, and shaped by
the nucleus's own accretion and feedback history, so the model captures it
only through a smooth, spherically-averaged radial function. Departures from
that idealisation -- a different effective slope, density fluctuations, or
large-scale asymmetry -- feed straight into the swept-mass history, deceleration profile, and hence
shift the crossing time. Because the crossing
depends on the \emph{differential} deceleration of the two shocks and the
sign of any departure is not fixed a priori, we regard the structure of the
circumnuclear medium as a moderate astrophysical uncertainty in $t_{\rm coll}$.

The outflow solid angles are a further systematic, and they couple to the
inferred speeds. Since $M_{\rm sw}\propto\Omega$, a more collimated outflow
sweeps less mass per unit solid angle, decelerates less, and is inferred to
be correspondingly faster at fixed energy. We fix the wind to
$\Omega_1=4\pi$, which fixes a particular $\beta_{\rm wind}$ and
$R_{\rm wind}(t)$; if the prompt outflow is even mildly aspherical, the wind
radius history, and the crossing, shift to later times as the wind is moving faster. The jet solid angle $\Omega_2$ is
fit, but the same $\Omega$--speed degeneracy applies, and restricting the
posterior to more or less collimated jets changes both the inferred dynamics and the
collision probability (the fits do not particularly well constrain $\Omega$).

The mass-injection history enters the dynamics through
$\epsilon\equiv M_{\rm sw}/\Delta M_{\rm caught}$, the ratio of swept-up to
caught-up mass that sets the deceleration, so the temporal shape of
$\dot M(t)$ --- and in particular \emph{when} the bulk of the injected mass
is delivered to the front --- governs the coast-to-decelerate transition
and the curvature of $R(t)$ near the crossing. We adopt a shifted power-law
engine with a single late-time slope $q$, the simplest member of a family
of injection profiles. A different functional form (for instance a broken
power law) distributes the mass differently and yields a
different deceleration, so the engine parametrisation is itself a source of
systematic uncertainty. The injection history is constrained by the rise
and decline of each flare, but it trades off against the density
normalisation and the injection velocity, and that degeneracy is inherited
by $t_{\rm coll}$. The systematic is most severe when the collision is predicted to be within the engine dominated regime 
$t_{\rm col}\sim {\rm few} \times t_{\rm start}$, which the AT2024tvd jet model fit prefers. 
The same
injection window also fixes for how long the trailing ejecta continues to
catch and re-energise the leading shock through internal shocks, which
further couples the injection history to the inferred velocity, and hence to
the predicted collision time.

Two effects bias the \emph{definition} of the crossing, with opposite
signs. The condition $R_{\rm wind}=R_{\rm jet}$ refers to the leading
shocks, whereas the jet reaches the back of the finite-width wind shell at
$R_{\rm jet}=R_{\rm wind}-\Delta R_{\rm wind}$, so the interaction
\emph{begins} slightly earlier than the quoted $t_{\rm coll}$, by
$\sim\Delta R_{\rm wind}/(\beta_{\rm rel}c)$ with $\Delta R_{\rm wind}\sim
R/{\rm few}$. Conversely, the front-crossing precedes the \emph{observable}
re-brightening: the transmitted shock must develop, $\nu_a$ must climb into
the band, and the declining second flare must fall below the new emission,
so the third flare emerges \emph{after} $t_{\rm coll}$. A reader should
therefore not expect the flare exactly at the quoted crossing time (for AT2020vwl we observe a third flare start time between 1225--1334\,d compared to a predicted crossing time of 1138$^{+275}_{-126}$\,d).

The start times carry their own systematic. Because $R(t)=\int_{t_{\rm
start}}^{t}\beta c\,{\rm d}t'$, the radius zero-point is $t_{\rm start}$.
The delay between flares $t_{\rm start,2}-t_{\rm start,1}$ directly sets the head-start
the wind has over the jet. Both start times are fit from the early light
curves under the assumption that the model accurately describes the very early evolution. Any unmodeled early-time physics (for example free-free
absorption from an external ionised screen, which would suppress the early,
low-frequency flux) would be absorbed into a shifted $t_{\rm start}$ and
propagate directly into $t_{\rm coll}$. As an explicit example of this last effect, \citet{Sfaradi25} modeled the AT2024tvd flare as being launched at $t=0$ with free-free absorption present, finding a reasonable fit. Our forward model can describe the flare evolution with a somewhat delayed ($t_{\rm start, 1}\approx 80$ d) launch time, and purely synchrotron self absorption. This slight delay in launch time  is entirely consistent with when $\dot m>1$ is reached for AT2024tvd (as seen in disk model fitting, \citealt{Guolo26}), but in principle an 80 day start time ambiguity would propagate directly into a collision time ambiguity. 

Finally, we have made various standard simplifying assumptions in the model framework, for example the existence of simple constant efficiency factors $\varepsilon_e, \varepsilon_B$ (which may well in reality vary with time and shock front physics), and a single non-thermal electron population described by a single (time independent) index $p$. Deviations from all of these simplifying assumptions are of course expected in reality. 

Owing to these various systematics, we are not (yet) concerned about the lack of an observed third flare from AT2024tvd, as one could play a model systematic game and shift the predicted collision time (we choose not to do this so that all modeling is consistent between sources).  We predict merely that a third flare should be imminent in AT2024tvd.

\subsection{Which TDEs are likely to show third flares?}
\begin{figure*}
    \centering
    \includegraphics[width=0.7\linewidth]{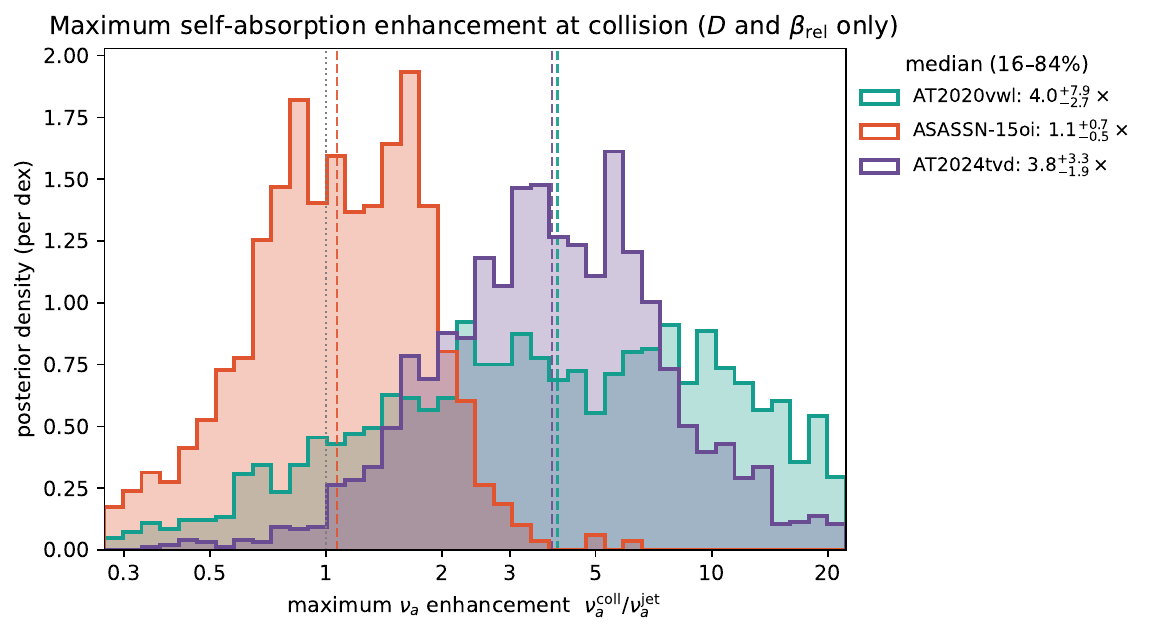}
    \caption{The distribution of the maximum relative shift in the synchrotron self absorption frequency as a result of the collisions between the forward modeled jet and winds for ASASSN-15oi, AT2020vwl and AT2024tvd, for the MCMC samples of the light curve fits in this work. The maximum $\nu_a$ enhancement is controlled almost entirely by the density contrast $D = n'_{\rm wind}/n_{\rm jet}$, with the relative velocity suppression factor $\beta_{\rm rel}/\beta_{\rm jet}$ impacting the relatively slowly moving (latest time) jet in ASASSN-15oi. The density contrast in AT2020vwl is more poorly constrained than the other two sources, but predicts an increase in $\nu_a$. ASASSN-15oi is predicted to have the smallest relative change, with AT2020vwl higher (though more uncertain). AT2024tvd is predicted to have a more reliably larger relative change, and as such is predicted to have a stronger third flare (which should appear imminently). The true change in synchrotron self absorption frequency also picks up a $\Delta R_{\rm coll}/\Delta R_{\rm jet}$ suppression factor.  }
    \label{fig:collision}
\end{figure*}

In the framework we present, it is clear that not all TDEs are likely to show detectable third radio flares. A strong wind-jet collision requires:
\begin{itemize}
    \item \textbf{A strong super-Eddington wind.} Not all TDEs are expected to reach highly super-Eddington accretion rates at early times (indeed, population synthesis disk modeling only predicts $\sim 40\%$ of TDEs should, \citealt{GoodwinMummery26}), confirmed by the fact that observationally not all TDEs show prompt radio emission \citep[e.g.][]{Alexander20}. It helps if a reasonable fraction of a solar mass is launched into such a wind, so that the wind is still actively shocking the CNM when the jet catches up (so that the jet notices a density increase).
    \item \textbf{A jet launched soon enough after the wind.} Delayed jets are launched at a critical accretion rate of $\approx2\%$ Eddington, and the time taken for the accretion disk to reach this critical accretion rate for the state transition to occur is highly variable across different TDE systems. Indeed, in the modeling of \citealt{GoodwinMummery26}, the predicted state transition time of a sampled TDE population includes sources for which $\Delta t > 10^4$ days, by which time even a relatively slowly moving wind will have traveled a long way, or conversely decelerated completely (so that there is not a shock front for the jet to collide with). 
    \item \textbf{Environment}. Extremely dense environments will decelerate the wind/jet more quickly, so that what it means for the wind to still be propagating ``strongly'' when the jet catches up is dependent on the ambient medium. 
\end{itemize}

Clearly, not all outflows launched from TDEs will provide conditions favourable for a third flare to be observed. Indeed, we show that the wind-jet collision for ASASSN-15oi will likely have minimal consequences on the radio emission from this source, as by the time the collision is predicted to occur both the wind and jet are moving at comparable speeds (so that the relative velocity is low) with density contrast which is insufficient to make up for this factor. 

To be quantitative, we compute the posterior predictions of 
\begin{equation}
    \left.\frac{\nu_a^{\rm coll}}{\nu_a^{\rm jet}}\right|_{\rm max} \;=\;
    D^{\frac{p+6}{2(p+4)}}
    \left(\frac{\beta_{\rm rel}}{\beta_{\rm jet}}\right)^{\frac{p+2}{p+4}} ,
\end{equation}
where $D \equiv {n'_{\rm wind}/ n_{\rm jet}}$ is the density contrast experienced by the jet relative to its own pre-collision ambient density, for each of the three TDEs modeled in this work at the time at which the shock fronts are predicted to collide (Figure \ref{fig:collision}). ASASSN-15oi has a jet which barely catches up with the wind, and is not predicted to do anything particularly interesting. AT2020vwl is more poorly constrained, but is entirely consistent with the observed $\sim$ doubling of the synchrotron self absorption frequency beyond the predicted collision time (consistent with the observed cahnged in $\nu_a$ in Figure \ref{fig:radio_lc}). AT2024tvd is predicted to have a more reliably larger relative change, and as such is predicted to have a stronger third flare, which should appear imminently, subject to systematic uncertainties discussed above.  

There are three other double flare sources currently known in the TDE literature, namely AT2019ahk \citep{Christy24}, AT2019dsg \citep{Cendes24} and AT2020zso \citep{Christy26}. Both AT2019dsg and AT2019ahk have relatively poorly constrained second flares, either due to minimal spectral coverage or simply because the second flare was relatively weak. AT2020zso is more amenable to the type of analysis performed here, as it shows two unambiguous radio flares, with clearly distinct properties (and well separated radial scales, \citealt{Christy26}).  AT2020zso in particular, and all of these sources, are well worth following up at late times in the radio.

\newpage
\section{Conclusions}\label{conc}
We have developed and then tested a specific prediction of the framework in which TDE radio
flares are produced by two distinct outflows (an early super-Eddington wind
and a later, faster, jet). If a single source launches both, the jet must
eventually overtake the wind, and the density jump it meets at the
swept-up wind shell may drive a third radio flare. Forward-modelling the
first two flares of AT2020vwl predicts the two shock fronts to cross at
$t_{\rm coll}$ (Fig.~\ref{fig:cross_vwl}), and new observations reveal a
third flare starting consistent with that time. To our knowledge this is the first TDE radio flare predicted in advance from the dynamics of earlier flares.

The physical mechanism by which this third flare originates is a simple self-absorption frequency jump.  When the jet shock enters
the dense, shocked wind shell, its self-absorption frequency $\nu_a$ and
self-absorbed peak flux rise, re-brightening the source with no change in
the central engine. The size of the shift is governed by three quantities:
the density contrast $D=n_{\rm wind}'/n_{\rm jet}$, the relative velocity $\beta_{\rm rel}$, and the
collision-layer width $\Delta R_{\rm coll}$ (Eq.~\ref{eq:nua_factors}). The simplest (possibly naive) model of the synchrotron self absorption frequency jump is ${\nu_a^{\rm coll}}/{\nu_a^{\rm jet}} \sim [({\gamma_{\rm ad}+1)/(\gamma_{\rm ad}-1})]^{{(p+6)}/{(2p+8)}} \sim 2.5$, where $\gamma_{\rm ad}=5/3$ is the adiabatic index of the circumnuclear gas. 

Applied to two further double-flare TDEs, the same modelling yields two
forward predictions. ASASSN-15oi has a small inferred density contrast and
a slow, late collision, so we expect at most a weak change in its radio
properties rather than a strong third flare, consistent with the absence
of a clear flare in the current data. A slight break in the evolution of $\nu_a$ for ASASSN-15oi has been observed (Fig. \ref{fig:vas_15oi_tvd}), but it is inconclusive with the sparse light curve sampling.  AT2024tvd, by contrast, has a fast
jet and the largest predicted self-absorption enhancement of the three, and
its shock fronts are predicted to cross near the end of the present observational coverage; a third flare should therefore emerge shortly afterwards, with the re-brightening pushed to higher frequencies. Continued radio monitoring provides a direct test of this framework.

Taken together, AT2020vwl (confirmed), ASASSN-15oi (weak), and
AT2024tvd (imminent, if detected) suggest that third flares are a generic, and
predictable, feature of TDEs that show both a prompt and a delayed flare.
They offer an independent test of the wind-plus-jet picture and a probe of
the circumnuclear density structure carved by the earlier outflow. A
quantitative treatment of the post-collision emission (including the transmitted
and reflected shocks, and the development of the emitting layer) is
beyond the scope of the present forward model and is left to future work,
as is the application of this framework to the growing sample of
multiply-flaring radio TDEs.

\newpage
\section*{Acknowledgments}
A.M. is grateful to Clement Bonnerot, Ryan Chornock and Raffaella Margutti for comments and discussions. 
A.M. acknowledges support from the Ambrose Monell Foundation, the W.M. Keck Foundation and the John N. Bahcall Fellowship Fund at the Institute for Advanced Study. A.J.G. is grateful for support from the Forrest Research Foundation. 
The National Radio Astronomy Observatory and Green Bank Observatory are facilities of the U.S. National Science Foundation operated under cooperative agreement by Associated Universities, Inc.
We thank the staff of the GMRT that made these observations possible. GMRT is run by the National Centre for Radio Astrophysics of the Tata Institute of Fundamental Research.

\bibliography{andy}
\bibliographystyle{aasjournal}

\begin{appendix}

\section{Radio data reduction} \label{sec:radio_dat_reduction}
A summary of the new radio observations presented in this work is given in Table \ref{tab:radio_obs}. In the following sections we describe the radio data reduction for each instrument. 

\subsection{VLA}
We observed the coordinates of AT2020vwl on 6 occasions with the VLA between 2024-07-02 and 2026-04-17 as part of our ongoing long-term monitoring program (Program IDs: 26A-167, 25B-154, 25A-159, 24B-108, 24A-159, 20B-377; PI: Goodwin, Christy, Alexander). All data were reduced following standard procedures in the Common Astronomy Software Package \citep[CASA;][]{Casa2022}, including the VLA calibration pipeline. We used 3C386 for flux and bandpass calibration in all observations and for phase calibration we used ICRF
J151340.1+233835 at 2--15\,GHz (Ku-, X-, C-, and S-
band); and ICRF J160207.2+332653 for 1--2 GHz (L-
band). Observations were carried out at a central frequency of 1.5 GHz (L-band) with 1 GHz of bandwidth, 3 GHz (S-band) with 2 GHz of bandwidth, 6 GHz (C-band) with 4 GHz of bandwidth, 10 GHz (X-band) with 4 GHz of bandwidth, and 15 GHz (Ku-band) with 6 GHz of bandwidth,  with each observation utilizing some subset of the receivers. 8-bit samplers were used for L- and S-band and 3-bit samplers were used for Ku-, X-, and C- bands.
We used the CASA task \texttt{tclean} to create images of the target field after a small amount of manual radio frequency interference (RFI) flagging, splitting the data into 0.5 or 1\,GHz sub-bands where sufficient SNR was achieved to make this possible. In all observations, AT2020vwl was detected as a point source. We measured the flux density of the point source using the CASA task \texttt{imfit} by fitting a 2D-Gaussian fixed to the size of the synthesized beam. We include both the statistical fit error on the reported flux densities plus a systematic 5$\%$ error to account for the accuracy of the absolute flux density calibration, added in quadrature.

\subsection{GMRT}
We observed the coordinates of AT2020vwl on three occasions with the GMRT at central frequencies of 0.65\,GHz (band 4) with a bandwidth of 300\,MHz and 1.26\,GHz (band 5) with a bandwidth of 400\,MHz as part of our ongoing long-term monitoring program (program IDs: 50$\_$040, 49$\_$078, and 48$\_$056, PI: Goodwin). Data reduction was performed manually in CASA using the same procedures outlined in \citet{Goodwin2023}. In all observations, 3C286 was used for flux and bandpass calibration and ICRF J160207.2+332653 for phase calibration. Images of the target field were created using the CASA task \texttt{tclean}, where we imaged the full band for each observation. The band 4 observations in December 2025 and February 2025 suffered severe RFI issues and a science quality image was not able to be produced. We exclude these data from our work AT2020vwl was detected as a point source in each of the other images. As for the VLA observations, we measured the flux density of the source by fitting an elliptical Gaussian fixed to the size of the synthesized beam using the CASA task \texttt{imfit}. We include both the statistical fit error on the reported flux densities plus a systematic 5$\%$ error to account for the accuracy of the absolute flux density calibration, added in quadrature.

\begin{longtable}{ccccc}
    \centering
Date & Time (d)$^*$ & Instrument & Frequency (GHz) & Flux density ($\mu$Jy) \\
\hline
\hline
2024-07-13 & 1334 & VLA  & 1.14  & $865 \pm 85$ \\
2024-07-13 & 1334 & VLA  & 1.35  & $774 \pm 41$ \\
2024-07-13 & 1334 & VLA  & 1.65  & $782 \pm 69$ \\
2024-07-13 & 1334 & VLA  & 1.9   & $593 \pm 36$ \\
2024-07-13 & 1334 & VLA  & 2.24  & $561 \pm 55$ \\
2024-07-13 & 1334 & VLA  & 2.75  & $561 \pm 30$ \\
2024-07-13 & 1334 & VLA  & 3.25  & $332 \pm 26$ \\
2024-07-13 & 1334 & VLA  & 3.76  & $325 \pm 25$ \\
2024-07-13 & 1334 & VLA  & 4.49  & $292 \pm 20$ \\
2024-07-13 & 1334 & VLA  & 5.51  & $227 \pm 23$ \\
2024-07-13 & 1334 & VLA  & 7     & $180 \pm 18$ \\
2024-07-13 & 1334 & VLA  & 9     & $136 \pm 12$ \\
2024-07-13 & 1334 & VLA  & 11    & $101 \pm 15$ \\
2024-07-13 & 1334 & VLA  & 15.08 & $56 \pm 6$ \\
\hline
2024-09-11 & 1394 & VLA  & 1.14  & $747 \pm 51$ \\
2024-09-11 & 1394 & VLA  & 1.39  & $749 \pm 32$ \\
2024-09-11 & 1394 & VLA  & 1.65  & $612 \pm 140$ \\
2024-09-11 & 1394 & VLA  & 1.9   & $478 \pm 40$ \\
2024-09-11 & 1394 & VLA  & 2.25  & $483 \pm 44$ \\
2024-09-11 & 1394 & VLA  & 2.75  & $352 \pm 31$ \\
2024-09-11 & 1394 & VLA  & 3.25  & $332 \pm 25$ \\
2024-09-11 & 1394 & VLA  & 3.75  & $328 \pm 27$ \\
2024-09-11 & 1394 & VLA  & 4.49  & $236 \pm 20$ \\
2024-09-11 & 1394 & VLA  & 5.51  & $182 \pm 26$ \\
2024-09-11 & 1394 & VLA  & 7.25  & $142 \pm 17$ \\
2024-09-11 & 1394 & VLA  & 9     & $124 \pm 14$ \\
2024-09-11 & 1394 & VLA  & 11    & $109 \pm 17$ \\
2024-09-11 & 1394 & VLA  & 13.55 & $68 \pm 9$ \\
2024-09-11 & 1394 & VLA  & 16.62 & $53 \pm 10$ \\
\hline
2025-02-07 & 1543 & VLA  & 1.5   & $408 \pm 30$ \\
2025-02-07 & 1543 & VLA  & 2.24  & $443 \pm 50$ \\
2025-02-07 & 1543 & VLA  & 2.75  & $268 \pm 29$ \\
2025-02-07 & 1543 & VLA  & 3.24  & $319 \pm 21$ \\
2025-02-07 & 1543 & VLA  & 3.75  & $212 \pm 21$ \\
2025-02-07 & 1543 & VLA  & 4.49  & $204 \pm 25$ \\
2025-02-07 & 1543 & VLA  & 5.51  & $102 \pm 29$ \\
2025-02-07 & 1543 & VLA  & 7     & $158 \pm 29$ \\
2025-02-07 & 1543 & VLA  & 9     & $93 \pm 15$ \\
2025-02-07 & 1543 & VLA  & 11    & $97 \pm 19$ \\
2025-02-13 & 1549 & GMRT & 1.26  & $543 \pm 20$ \\
\hline
2025-07-04 & 1690 & VLA  & 1.14  & $710 \pm 204$ \\
2025-07-04 & 1690 & VLA  & 1.39  & $444 \pm 60$ \\
2025-07-04 & 1690 & VLA  & 1.65  & $474 \pm 106$ \\
2025-07-04 & 1690 & VLA  & 1.9   & $211 \pm 45$ \\
2025-07-04 & 1690 & VLA  & 2.24  & $303 \pm 56$ \\
2025-07-04 & 1690 & VLA  & 2.76  & $256 \pm 30$ \\
2025-07-04 & 1690 & VLA  & 3.24  & $227 \pm 25$ \\
2025-07-04 & 1690 & VLA  & 3.76  & $185 \pm 24$ \\
2025-07-04 & 1690 & VLA  & 4.49  & $148 \pm 20$ \\
2025-07-04 & 1690 & VLA  & 5.51  & $105 \pm 21$ \\
2025-07-04 & 1690 & VLA  & 6.49  & $96 \pm 18$ \\
2025-07-04 & 1690 & VLA  & 7.51  & $113 \pm 18$ \\
2025-07-04 & 1690 & VLA  & 9     & $73 \pm 12$ \\
2025-07-04 & 1690 & VLA  & 11    & $36 \pm 13$ \\
2025-08-01 & 1718 & GMRT & 0.65  & $1122 \pm 25$ \\
2025-08-01 & 1718 & GMRT & 1.26  & $505 \pm 59$ \\
\hline
2025-12-02 & 1841 & VLA  & 1.14  & $442 \pm 97$ \\
2025-12-02 & 1841 & VLA  & 1.39  & $317 \pm 36$ \\
2025-12-02 & 1841 & VLA  & 1.65  & $458 \pm 71$ \\
2025-12-02 & 1841 & VLA  & 1.9   & $299 \pm 42$ \\
2025-12-02 & 1841 & VLA  & 9     & $37 \pm 15$ \\
2025-12-02 & 1841 & VLA  & 11    & $33 \pm 16$ \\
2025-12-02 & 1841 & GMRT & 1.26  & $397 \pm 21$ \\
\hline
2026-04-17 & 1977 & VLA & 1.5 & 234$\pm16$ \\
2026-04-17 & 1977 & VLA & 10 & 27$\pm$9\\
\hline
\hline
    \caption{New radio observations of AT2020vwl presented in this work. $^*$All times are measured from optical peak, MJD 59170.}
    \label{tab:radio_obs}
\end{longtable}

\section{Radio spectral fitting of individual epochs}\label{app:spec}

\begin{figure*}
    \centering
    \includegraphics[width=0.5\linewidth]{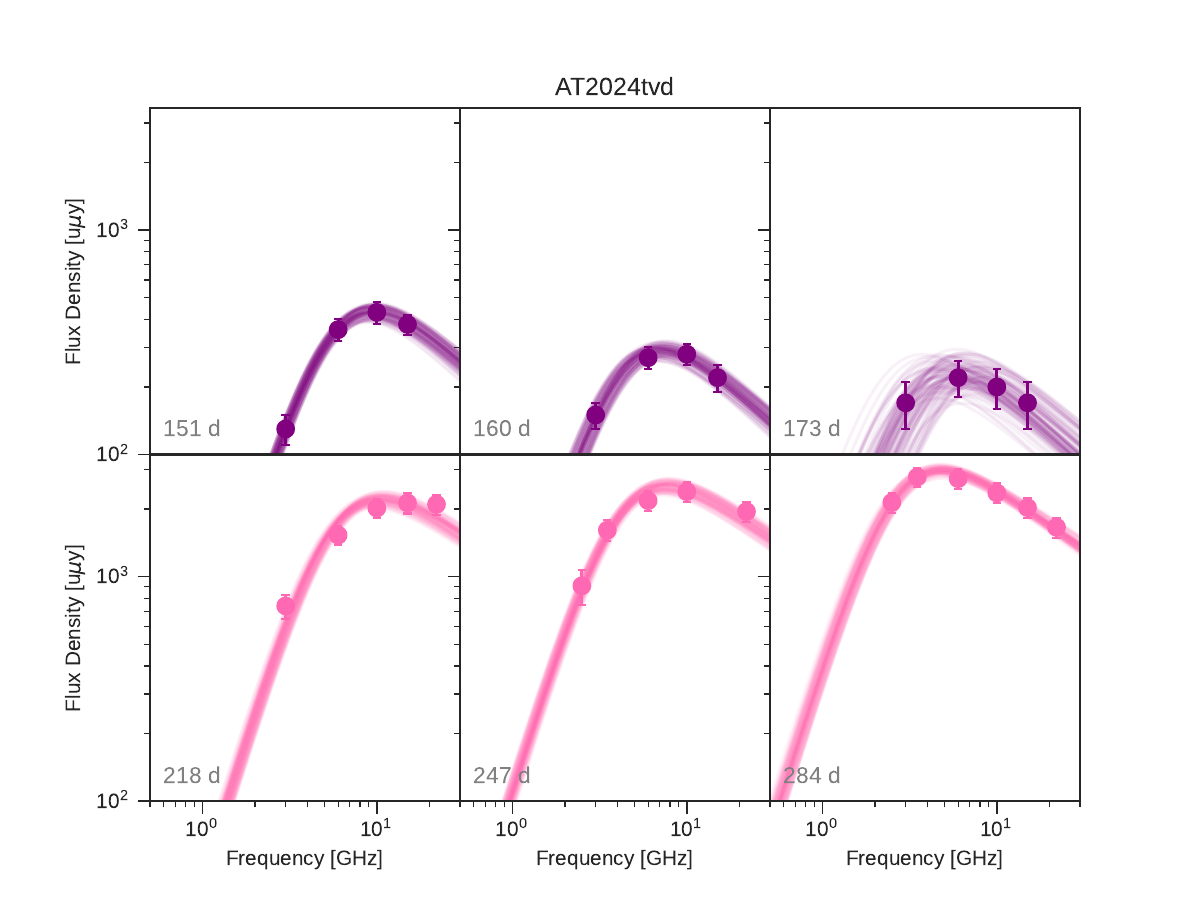}
    \includegraphics[width=0.49\linewidth]{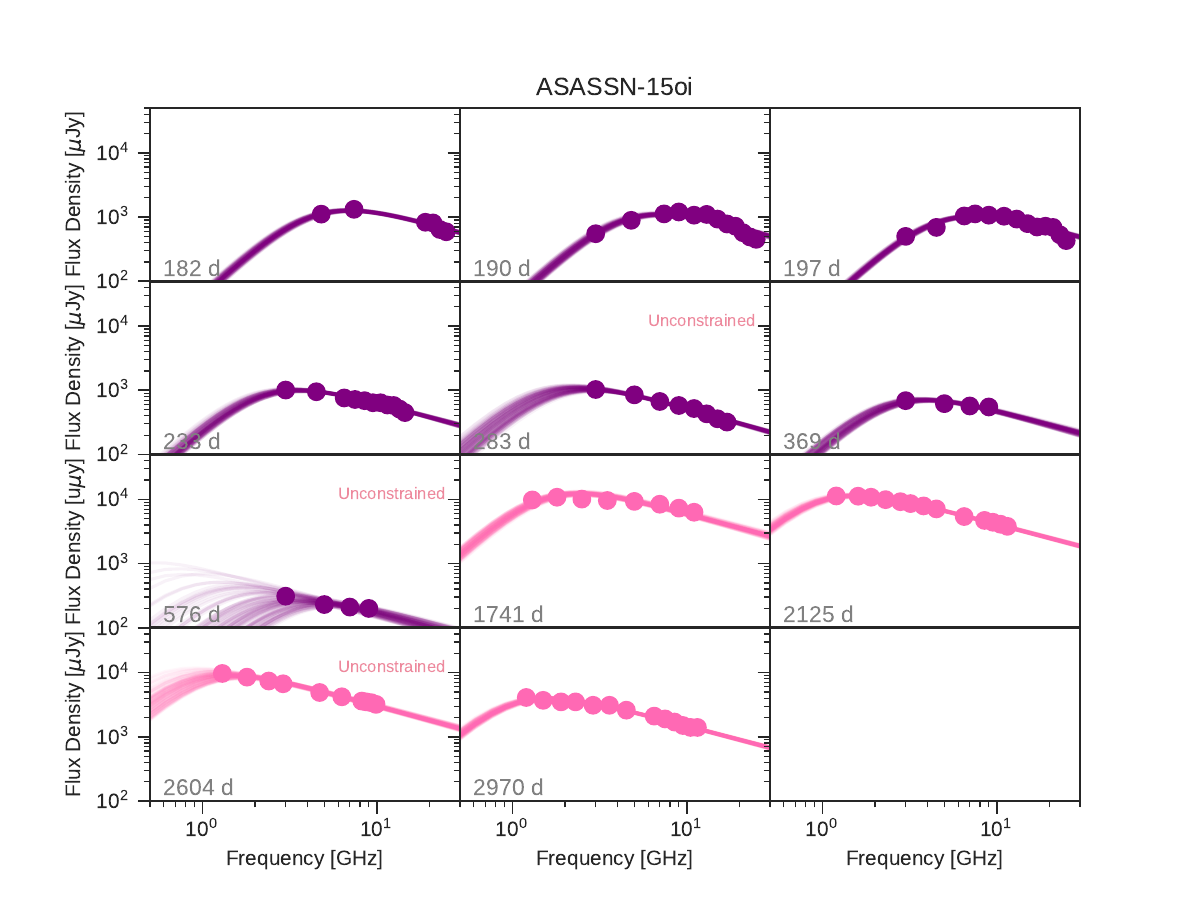}
    \includegraphics[width=0.69\linewidth]{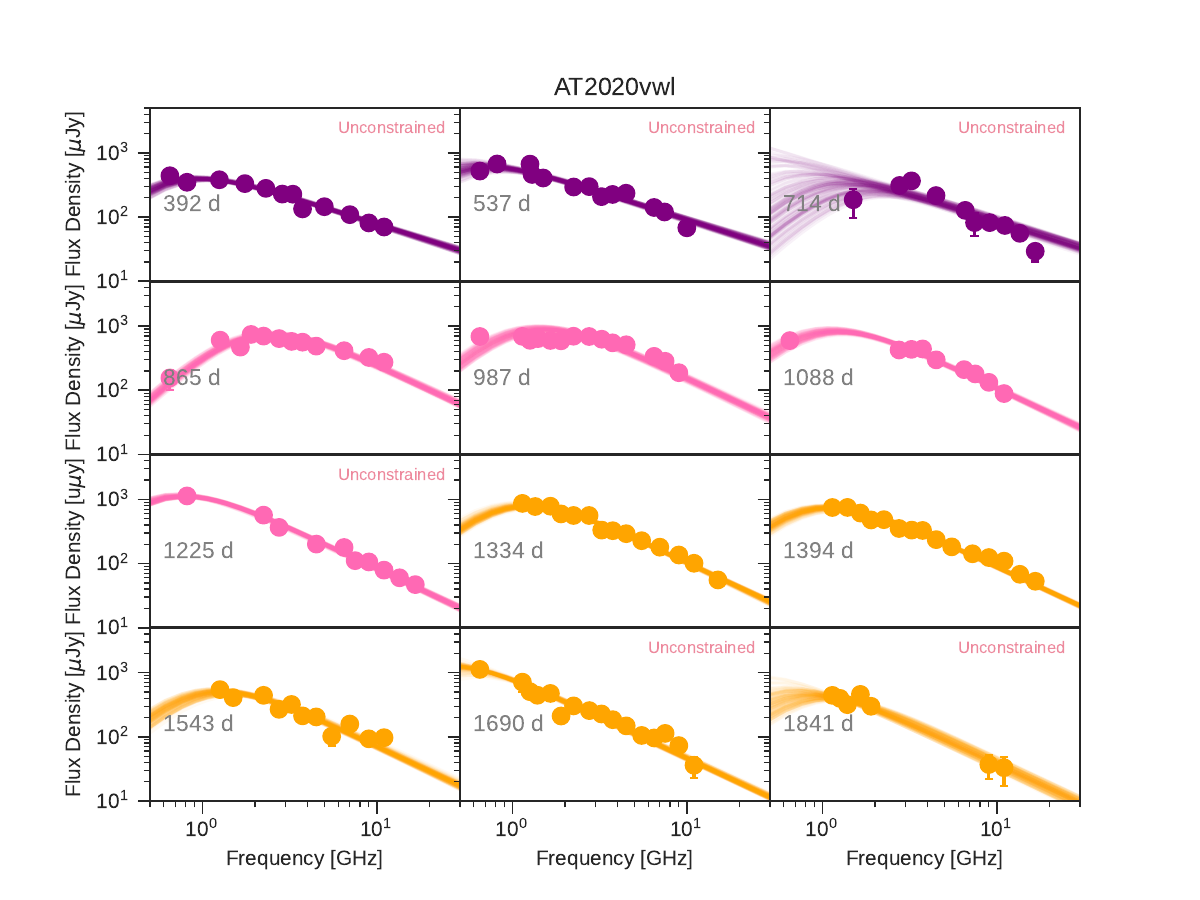}
    
    \caption{A selection of spectral fits to the radio observations of AT2020vwl (top left), ASASSN-15oi (top right), and AT2024tvd (bottom). Spectra are coloured by first flare (purple), second flare (pink), and third flare (orange). We determine if $\nu_a$ is constrained by requiring the following criteria is satisfied: $\nu_a$ is constrained to be within the range of frequencies the observations cover ($>0.65$\,GHz) and $\nu_a$ is well-constrained in the MCMC posterior distribution at $>10\sigma$ significance. Panels labeled ``unconstrained" mean that the constraint on $\nu_a$ did not reach these criteria. 
    }
    \label{fig:spec_fits}
\end{figure*}

In Figure \ref{fig:spec_fits} we show the spectral fits described in Section \ref{sec:spectra} as well as the spectral fits for ASASSN-15oi and AT2024tvd, using the same fitting approach as described in Section \ref{sec:spectra}. We fit only one spectrum for each epoch, but note that when there are two-emitting regions contributing to the observed spectrum, such as at 987\,d for AT2020vwl, this approach does not capture the peak of both spectra. Nevertheless the observations constrain that overall the peak of the spectrum for the second flare of AT2020vwl shifted to significantly higher frequencies than during the third flare, as is evident in Figure \ref{fig:radio_lc} right. It is also evident that ASASSN-15oi, despite having less-well constrained spectra, has not shown a significant increase in the self-absorption frequency since the onset of the second flare. 

\begin{figure}
    \centering
    \includegraphics[width=0.49\linewidth]{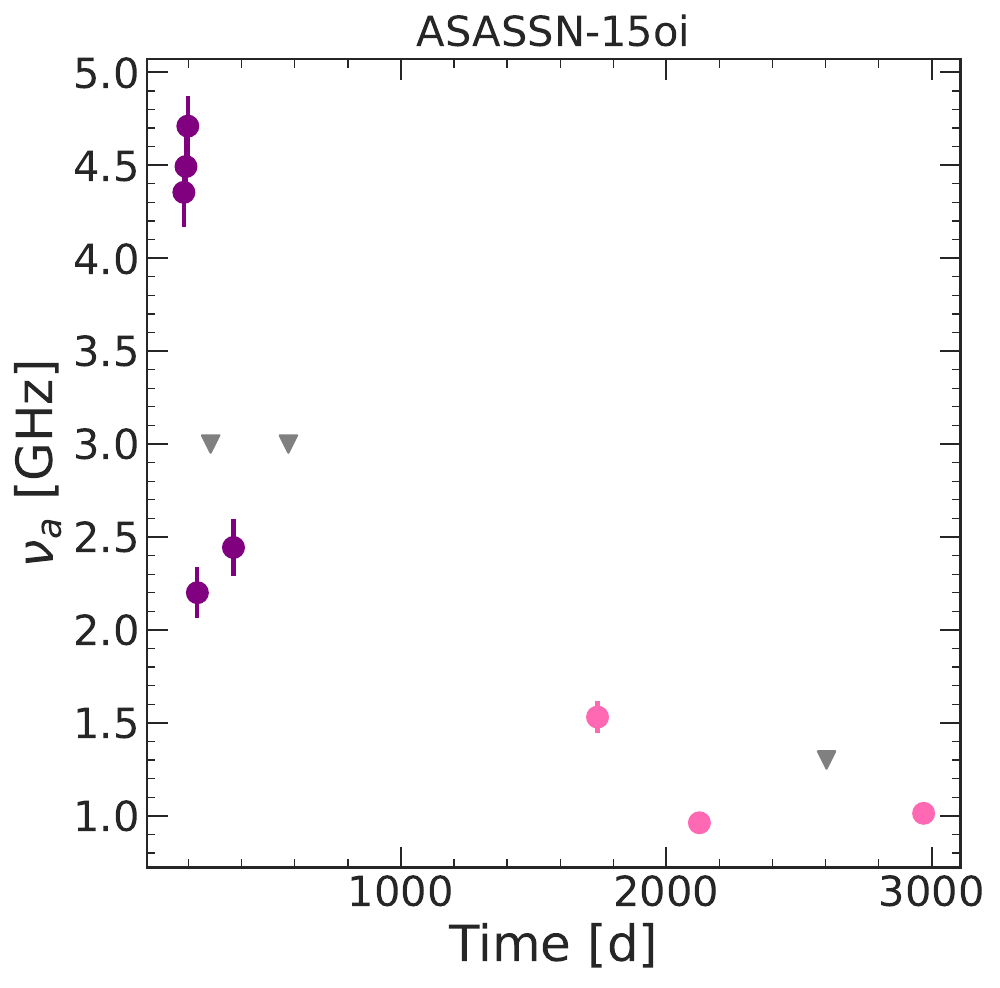}
    \includegraphics[width=0.47\linewidth]{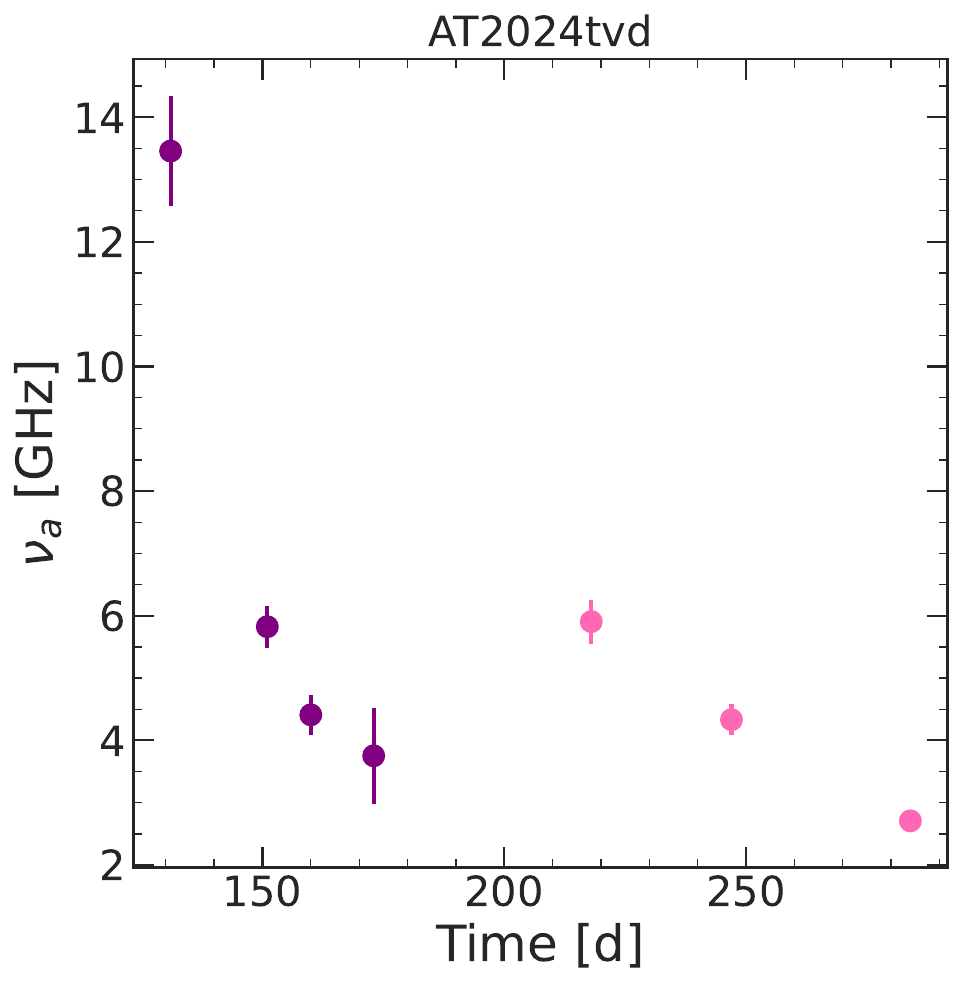}
    \caption{The evolution of the synchrotron self absorption frequency $\nu_a$ for ASASSN-15oi (left panel) and AT2024tvd (right panel). When poorly constrained we display upper limits by inverted triangles. During a flare the synchrotron self absorption frequency generally decays with time, as is observed in both sources  during the first flare and the initial stages of the second flare. ASASSN-15oi shows indications of a break in the temporal evolution of $\nu_a$ at late times (the final epoch), which is consistent with a change in density from a jet front catching up to a wind shell. The temporal sampling of the light curve and spectra are, however, insufficient to constrain this strongly. }
    \label{fig:vas_15oi_tvd}
\end{figure}

\subsection{Emitting region radius}

We infer the radius of the emitting region using the assumption of equipartition, as is standard for TDE outflow analysis \citep[e.g.][]{Alexander16,Cendes21,Goodwin22}. The radius of the outflow can be estimated from the observed spectrum peak flux density ($F_{\mathrm{\nu,a}}$), and peak frequency ($\nu_{\mathrm{a}}$), following the derivation in \citet{BarniolDuran2013}. We assume the radius is given by
\begin{equation}\label{eq:R_eq}
\begin{aligned}
\begin{split}
    R = \left({F_{\mathrm{\nu,a}} \over {\rm mJy}}\right)^{\frac{6+p}{13+2p}} \left(\frac{\nu_{\mathrm{a}}}{10\,\rm{GHz}}\right)^{-1}{\cal R_{\rm{synch}}}
\end{split}
\end{aligned}
\end{equation}
where we define the ``canonical" synchrotron radius,
\begin{equation}\label{eq:R_synch}
\begin{aligned}
\begin{split}
    {\cal R_{\rm{synch}}} = 1\times10^{17} \,\,(21.8 (525^{(p-1)})^{\frac{1}{13+2p}}
    \gamma_{m}^{\frac{2-p}{13+2p}}
     \left(\frac{d}{10^{28}\,\rm{cm}}\right)^{\frac{2(p+6)}{13+2p}}
    (1+z)^{-\frac{19+3p}{13+2p}}\\
    f_{\mathrm{A}}^{-\frac{5+p}{13+2p}} f_{\mathrm{V}}^{-\frac{1}{13+2p}} 4^{\frac{1}{13+2p}} \xi^{\frac{1}{13+2p}} 
    \epsilon^{\frac{1}{13 + 2p}}
    \quad\rm{cm}. 
\end{split}
\end{aligned}
\end{equation}
where $d$ is the luminosity distance to the source, $z$ is the redshift, $\xi = 1 + {1}/{\epsilon_{\rm e}}$, $\gamma_m = 2$ is the minimum Lorentz factor of the accelerated electrons, and $\epsilon = \frac{\epsilon_B}{\epsilon_e} \frac{11}{2(p+1)}$ parametrizes any deviation from equipartition. The factors $\epsilon_e$ and $\epsilon_B$ are the fraction of the thermal energy in the post-shocked medium which accelerates the electrons, and the fraction of the total energy in the magnetic field respectively. We use fiducial values of $\epsilon_e=0.1$ and $\epsilon_B=0.02$, and assume an outflow half-opening angle of 60\,deg, noting that here we are only interested in the relative change in the radius over time rather than the absolute value so the geometry and microphysics assumptions are unimportant. 

The inferred radii over time for each of the 3 flares of AT2020vwl is plotted in Figure \ref{fig:radius}. Evidently for AT2020vwl, there is no compact ``new" emitting region emerging during the third flare. This is distinct from what was observed at the onset of the second flare. 


\section{Parameters inferred}
\begin{figure}
    \centering
    \includegraphics[width=0.9\linewidth]{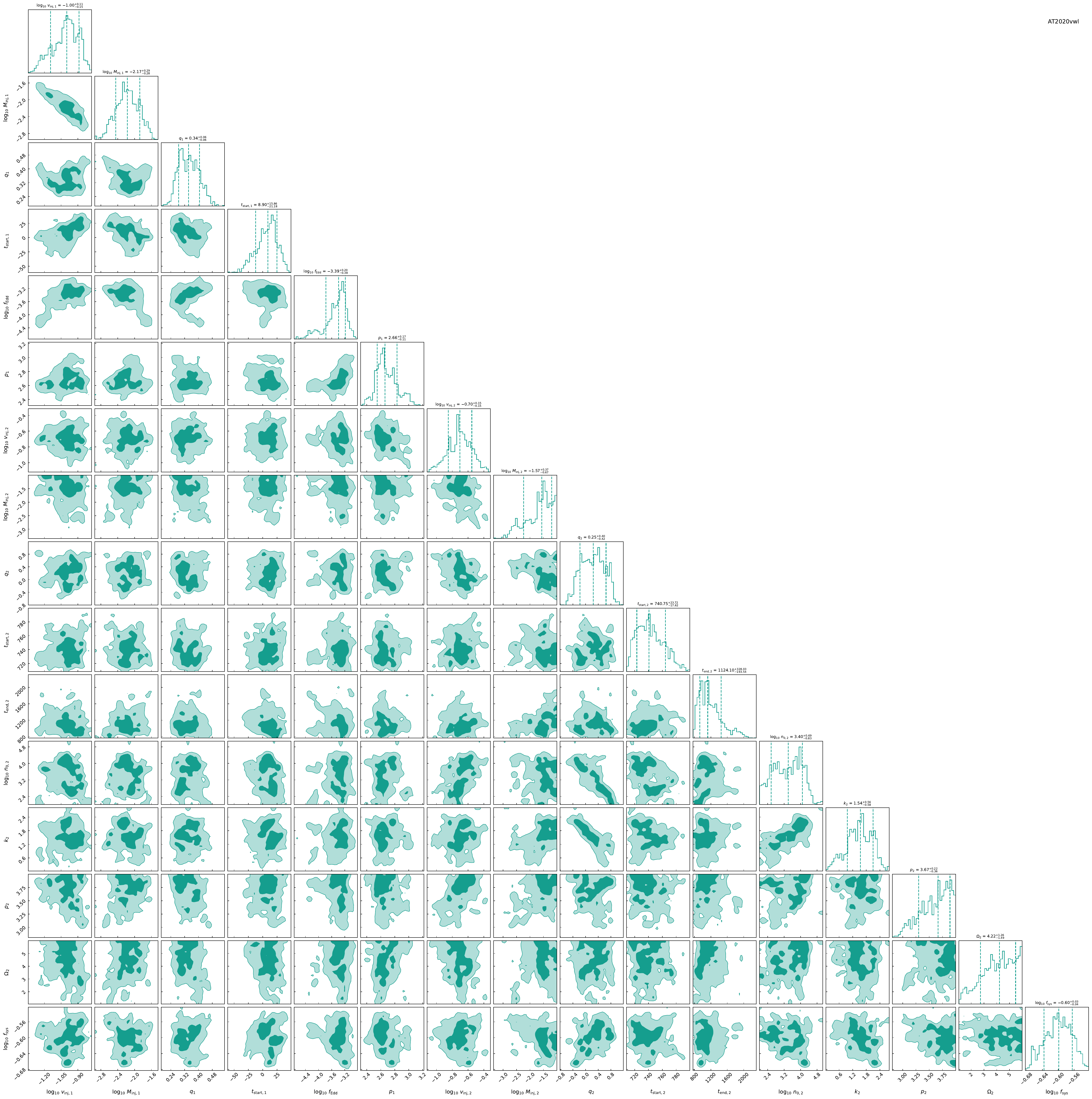}
    \caption{The posterior corner plot for the double flare fit to AT2020vwl. Parameters are in dimensionless code units, with velocities in units of $c$, masses in units of solar, densities in cgs units, and all times measured in days. }
    \label{fig:corner_20vwl}
\end{figure}

\begin{figure}
    \centering
    \includegraphics[width=0.9\linewidth]{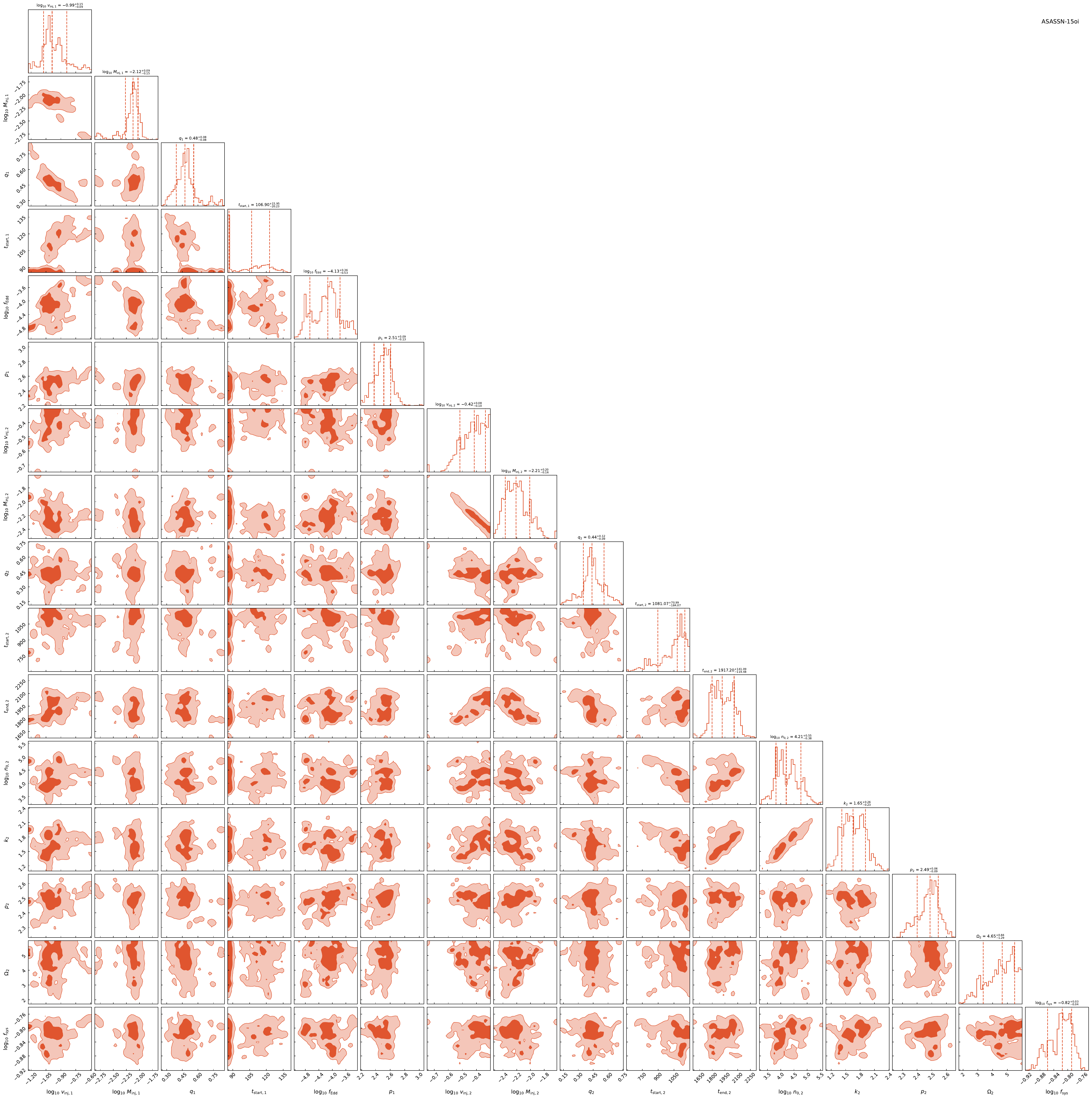}
    \caption{The posterior corner plot for the double flare fit to ASASSN-15oi. Parameters are in dimensionless code units, with velocities in units of $c$, masses in units of solar, densities in cgs units, and all times measured in days. }
    \label{fig:corner_15oi}
\end{figure}

\begin{figure}
    \centering
    \includegraphics[width=0.9\linewidth]{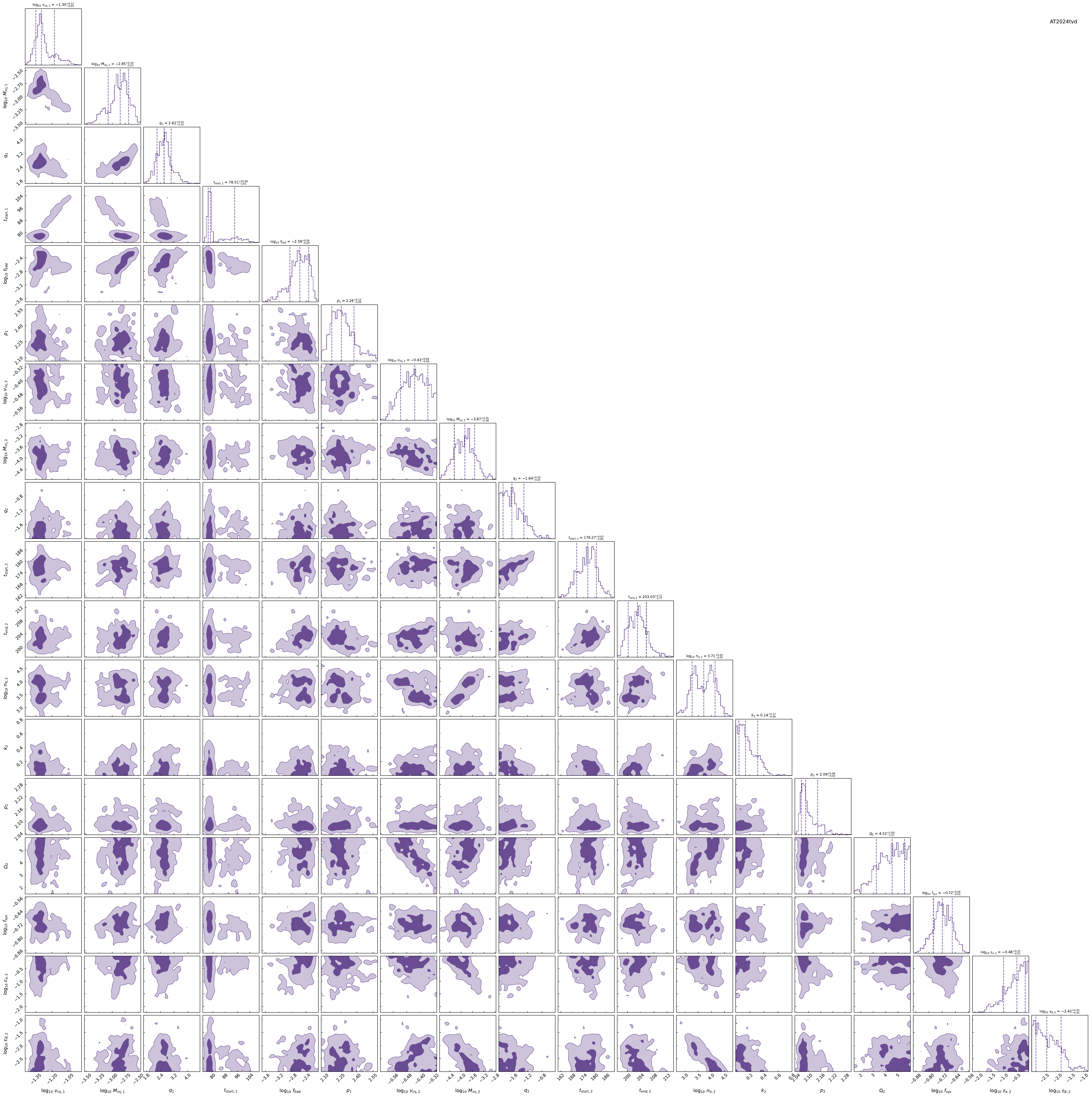}
    \caption{The posterior corner plot for the double flare fit to AT2024tvd. Parameters are in dimensionless code units, with velocities in units of $c$, masses in units of solar, densities in cgs units, and all times measured in days. As AT2024tvd shows evidence for a cooling break in its second flare, we leave the microphysical parameters $\varepsilon_e$ and $\varepsilon_B$ free for the second flare, in contrast with ASASSN-15oi and AT2020vwl.  }
    \label{fig:corner_24tvd}
\end{figure}

We show the three corner plots for our fits to AT2020vwl (Fig. \ref{fig:corner_20vwl}), ASASSN-15oi (Fig. \ref{fig:corner_15oi}) and AT2024tvd (Fig. \ref{fig:corner_24tvd}) in this Appendix. Parameters are in dimensionless code units, with velocities in units of $c$, masses in units of solar, densities in cgs units, and all times measured in days. As AT2024tvd shows evidence for a cooling break in its second flare, we leave the microphysical parameters $\varepsilon_e$ and $\varepsilon_B$ free for the second flare, in contrast with ASASSN-15oi and AT2020vwl.  

\end{appendix}

\end{document}